\documentclass[12pt]{article}
\usepackage{amsmath} 
\usepackage{pstricks}
\usepackage{longtable}
\usepackage[dvips]{graphicx}       
\usepackage{color}
\input{epsf} 
\newcommand\as{\alpha_{\mathrm{S}}} 
\newcommand\f[2]{\frac{#1}{#2}}

\def\beq{\begin{equation}} 
\def\eeq{\end{equation}} 
\def\to{\rightarrow} 
\def\nn{\nonumber}

\def\b0{\beta_0}

\def\GE{\gamma_E}
\def\beeq{\begin{eqnarray}}
\def\eeeq{\end{eqnarray}}
\def\ep{\epsilon}
\def\bom#1{{\mbox{\boldmath $#1$}}}
\def\cm{{\cal M}}
\def\soft{\overset{\underset{\mathrm{S}}{}}{\simeq}}

\newcommand\ao{\alpha_0}

\begin{document}

\begin{titlepage}
\renewcommand{\thefootnote}{\fnsymbol{footnote}}
\begin{flushright}
hep-ph/1209.0673\\
JHEP12(2012)088
\end{flushright}
\par \vspace{10mm}

\begin{center}
{\Large \bf
A next-to-next-to-leading order calculation \\[0.5cm]
of soft-virtual cross sections
}
\end{center}
\par \vspace{2mm}
\begin{center}
{\bf Daniel de Florian}\footnote{deflo@df.uba.ar} and
{\bf Javier Mazzitelli}\footnote{jmazzi@df.uba.ar}\\

\vspace{5mm}

Departamento de F\'\i sica, FCEyN, Universidad de Buenos Aires, \\
(1428) Pabell\'on 1, Ciudad Universitaria, Capital Federal, Argentina.\\

\vspace{5mm}

\end{center}

\par \vspace{2mm}
\begin{center} {\large \bf Abstract} \end{center}
\begin{quote}
\pretolerance 10000

We compute the next-to-next-to-leading order (NNLO) soft and virtual QCD corrections for the partonic cross section of colourless-final state processes in hadronic collisions.
The results are valid to all orders in the dimensional regularization parameter $\ep$. 
The dependence of the results on a particular process is given through finite contributions to the one and two-loop amplitudes, which have to be computed in a process-by-process basis. To evaluate the accuracy of the soft-virtual approximation we compare it with the full NNLO result for  Drell-Yan and Higgs boson production via gluon fusion. We also provide a universal expression for the hard coefficient needed to perform threshold resummation up to next-to-next-to-leading logarithmic (NNLL) accuracy.

\end{quote}

\vspace*{\fill}
\begin{flushleft}
December 2012
\end{flushleft}
\end{titlepage}

\setcounter{footnote}{1}
\renewcommand{\thefootnote}{\fnsymbol{footnote}}

\section{Introduction}

The development of accurate QCD calculations is a fundamental tool to properly test the Standard Model. 
Given the size of the perturbative corrections, leading order (LO) evaluations are insufficient, and higher perturbative orders must be taken into account.
However, these calculations are highly non-trivial, and at present only a few processes have been  computed analytically with full next-to-next-to-leading order (NNLO) precision. At hadronic colliders, only Drell-Yan \cite{vNeerven:fullDY,higgsNNLO1} and Higgs boson production \cite{higgsNNLO1,higgsNNLO2,Ravindran:2003um} (within the effective vertex approach) have reached that stage of accuracy.

Higgs production is a particular example of an observable with a slow convergence for the perturbative expansion in the strong coupling constant $\as$. Next-to-leading (NLO) corrections \cite{Dawson:1990zj,Djouadi:1991tk,Spira:1995rr} are as large as the Born result and the NNLO contribution still increases the cross section by about 25\% at LHC energies.
Analyses from scale variations \cite{deFlorian:2012yg,Baglio:2010ae,Anastasiou:2012hx,deFlorian:2009hc,Anastasiou:2008tj} and soft-gluon expansion \cite{Catani:2003zt,Moch:2005ky,Laenen:2005uz,Ravindran:2006cg,Kramer:1996iq} indicate that the next orders (N$^3$LO and beyond) can contribute still at the level of $10\%$. 

An important step towards a complete NNLO calculation for both Drell-Yan and Higgs production has been the evaluation of the soft and virtual contributions \cite{Matsuura:1988sm,Catani:2001ic,Harlander:2001is}, which provide the dominant terms for those processes.
As a matter of fact, this is a general feature when a system of large invariant mass $Q$ is produced in  hadronic collisions. 
Since parton distributions $f_{a/h}(x)$ grow very rapidly for small fractions  of the hadron momentum $x$, the partonic center-of-mass energy tends to be close to the invariant mass $Q$, and the remaining energy only allows for the emission of soft particles. For this reason the soft-virtual contributions are expected to be a very good approximation to the total cross section for a large number of processes.

In this paper we exploit the factorization properties of the QCD matrix elements to compute the soft and virtual contributions to the partonic cross sections at NNLO for a wide number of processes in hadronic collisions where a system of colourless particles is produced (as gauge bosons, Higgs, leptons, etc.). The computational approach presented here can be extended to higher orders in perturbation theory simplifying considerably the evaluation of the corrections.

 We present a {\it universal} expression for the corresponding cross section at NNLO, valid to all orders in the dimensional regularization parameter $\ep$.
With this result, it is possible to evaluate the soft-virtual approximation for any process of the kind studied in this paper in an automatized way once the 
relevant one- and two-loop amplitudes become available and, therefore, provide a first estimate of the size of higher order corrections for a number of interesting observables at the LHC.
Furthermore, since the results are valid to  all orders in $\ep$, they become necessary ingredients for ultraviolet and infrared factorization at N$^3$LO (and beyond) within the same approximation.

Another interesting use of the soft approximation is the relation to the soft-gluon threshold resummation approach. We profit from this calculation and obtain, for the first time, a {\it universal} expression for the hard coefficient needed to perform threshold resummation up to next-to-next-to-leading logarithmic (NNLL) accuracy.

The paper is organized as follows.
In section \ref{notacion} we present the notation for the QCD cross sections and show how phase space factorization occurs in the soft limit.
In sections \ref{nlo} and \ref{nnlo} we perform the calculation of the soft-virtual corrections at NLO and NNLO respectively.
In section \ref{sec:mellin} we present the soft-virtual approximation in Mellin space.
In section \ref{pheno} we analyse the phenomenological results for Drell-Yan and  Higgs boson production via gluon fusion to compare the soft-virtual approximation with the full result as a way to validate its dominance.
In section \ref{resum} we profit from the previous result and present a {\it universal} expression for the hard coefficient required to perform threshold resummation up to NNLL accuracy.
Finally, in section \ref{conc} we present our conclusions.

\section{QCD cross sections}\label{notacion}

We consider the following general process in hadronic collisions:
\beq\label{proceso}
h_1+h_2\longrightarrow F + X
\eeq
where $F$ denotes any colourless final state (i.e. without quarks or gluons), and $X$ stands for any inclusive final hadronic state. The center-of-mass energy is $\sqrt{s_H}$, and $Q$ is the invariant mass of the system $F$ which
 can involve a combination of  gauge bosons, Higgs, isolated photons, leptons, etc.
The inclusive cross section can be written as
\begin{align}
\label{had}
Q^2\,\f{d\sigma}{dQ^2}(s_H,Q^2) =& 
\sum_{a,b} \int_0^1 dx_1 \;dx_2 \; f_{a/h_1}(x_1,\mu_F^2) 
\;f_{b/h_2}(x_2,\mu_F^2) \int_0^1 dz \;\delta\!\left(z -
\frac{\tau}{x_1x_2}\right) \nn \\
& \times \hat{\sigma}_0\,z\;G_{ab}(z;\as(\mu_R^2), Q^2/\mu_R^2;Q^2/\mu_F^2) \;,
\end{align}
where $\tau=Q^2/s_H$, $\mu_F$ and $\mu_R$ are the factorization and 
renormalization scales respectively, and $\hat{\sigma}_0$ is the Born level partonic cross section. 
The parton densities of the colliding hadrons are denoted by 
$f_{a/h}(x,\mu_F^2)$ and the subscripts $a,b$ label the type
of massless partons ($a,b=g,q_f,{\bar q}_f$,
with $N_f$ different flavours of light quarks). 

According to Eq.~(\ref{had}), the cross section ${\hat \sigma}_{ab}$ for the partonic subprocess $ab \to F + X$ at the center-of-mass energy $s=x_1 x_2 s_H$ is
\begin{equation}\label{zfactor}
Q^2\,\f{d{\hat \sigma}_{ab}}{dQ^2}(s,Q^2) = \frac{1}{s} 
\;\hat{\sigma}_0\, Q^2 \;G_{ab}(z) = \hat{\sigma}_0 \;z \;\;G_{ab}(z) \;,
\end{equation}
where the term $1/s$ arises from the flux factor and leads to 
an overall $z$ factor, being $z=Q^2/s$ the partonic equivalent of $\tau$.
The hard coefficient function $G_{ab}$ has a perturbative expansion in terms of powers of the QCD renormalized coupling $\as(\mu_R^2)$:
\begin{align}
\label{expansion}
G_{ab}(z;\as(\mu_R^2), Q^2/\mu_R^2;Q^2/\mu_F^2) &=
\sum_{n=0}^{+\infty} \left(\f{\as(\mu_R^2)}{2\pi}\right)^n
\;G_{ab}^{(n)}(z;Q^2/\mu_R^2;Q^2/\mu_F^2)\;.
\end{align}
In the following, the dependence of $\as$ on the renormalization scale $\mu_R$ is understood.
We always use the $\overline{\text{MS}}$ scheme for the renormalization of the strong coupling.

At leading-order the partonic subprocess is $ab \to F$, and since the final state is colourless we can only have $ab=gg$ or $ab=q{\bar q}$ (and $ab={\bar q}q$). 
The LO contribution is then
\begin{equation}
G_{ab}^{(0)}(z) = \delta_{ag} \; \delta_{bg} \;\delta(1-z) \;
\;\;\;\;{\rm or}\;\;\;\;\;
G_{ab}^{(0)}(z) = (\delta_{aq} \; \delta_{b{\bar q}} + \delta_{a{\bar q}} \; \delta_{bq}) \;\delta(1-z)\;.
\end{equation}
At higher orders, other parton subprocesses can contribute to the total cross section.

In the soft-virtual approximation, however, we are only interested in the same parton subprocess present at LO, and we compute only those contributions in $G_{a\bar{a}}$ that give rise to the distributions $\delta(1-z)$ and ${\cal D}_i(z)$ in the coefficient function, where we have defined
\begin{equation}
{\cal D}_i(z) \equiv \left[ \f{\ln^i(1-z)}{1-z}\right]_+\,\, .
\end{equation}
The $+$ indicates the usual plus-prescription,
\beq
\int_0^1 {\cal D}_i(z) f(z)\, dz =
\int_0^1 \f{\ln^i(1-z)}{1-z}\left[\,f(z)-f(0)\right] dz\;.
\eeq 
These two types of contributions are the most singular terms when $z\rightarrow 1$, and then dominate the cross section in the soft limit.

The aim of this paper is to calculate the NNLO soft-virtual corrections for any process of the type of Eq.~(\ref{proceso}). The parton subprocesses that contribute up to second order in $\as$ are
\beeq\label{procesos}
\as^0\;\;\;&\;&\;\;a\bar{a}\to F \;\;\;\;\;\;\;\;\;\;\;\;\;\;\;\;\; \text{(tree-level)} \nn\\
\as^1\;\;\;&\;&\;\;a\bar{a}\to F \;\;\;\;\;\;\;\;\;\;\;\;\;\;\;\;\; \text{(one-loop)}\nn\\
\;\;\;&\;&\;\;a\bar{a}\to F + g \;\;\;\;\;\;\;\;\;\;\;
\text{(tree-level)} \nn\\
\as^2\;\;\;&\;&\;\;a\bar{a}\to F \;\;\;\;\;\;\;\;\;\;\;\;\;\;\;\;\; \text{(two-loop)}\\
\;\;\;&\;&\;\;a\bar{a}\to F + g \;\;\;\;\;\;\;\;\;\;\; \text{(one-loop)}\nn\\
\;\;\;&\;&\;\;a\bar{a}\to F + q + {\bar q}\;\;\;\;\;
\text{(tree-level)} \nn\\
\;\;\;&\;&\;\;a\bar{a}\to F + g + g \;\;\;\;\; \text{(tree-level)} \;\nn,
\eeeq
with $a\bar{a}=gg$ or $a\bar{a}=q{\bar q}$, depending on the process.

\subsection{Phase-space factorization}

To compute the real corrections of the processes we are interested in, we have to perform the phase-space integration of the corresponding matrix elements in the limit in which the emitted QCD-partons become soft. In this limit, the phase-space can be written in a factorized form.

Let us consider that the final state $F$ has $l$ non-QCD particles with momenta $\{k\}$, and $m$ soft QCD massless partons with momenta $\{q\}$. The $n$-dimensional phase-space is then
\beq\label{ps}
\int d\text{PS}=\int 
\left[\prod_{i=1}^{m}\frac{d^n q_i}{(2\pi)^{n-1}}\delta^+(q_i^2)\right]
\left[\prod_{j=1}^{l}\frac{d^n k_j}{(2\pi)^{n-1}}\delta^+(k_j^2-m_j^2)\right]
(2\pi)^n \delta^n(p_1+p_2-q_1-\,\dots\, q_m-k_1-\,\dots\, k_l)\,.
\eeq
We introduce the momentum $K=k_1+\dots k_l$ \cite{Ellis:1979sj}, with $K^2=Q^2$. Multiplying the above equation by the identity
\beq
1=\int d^nK\ \delta^n(K-k_1-\dots k_l) \int dQ^2\ \delta^+(K^2-Q^2)\;
\eeq
we arrive at
\beeq\label{ps2}
\int d\text{PS}=&&\!\!\!\!\!\!\!\!\!\f{dQ^2}{2\pi}\int 
\left[\prod_{j=1}^{l}\frac{d^n k_j}{(2\pi)^{n-1}}\delta^+(k_j^2)\right]
(2\pi)^n \delta^n(K-k_1-\,\dots\, k_l) \\
\times&&\!\!\!\!\!\!\!\!\!\int\left[\prod_{i=1}^{m}\frac{d^n q_i}{(2\pi)^{n-1}}\delta^+(q_i^2)\right]
\left[
\f{d^n K}{(2\pi)^{n-1}}\delta^+(K^2-Q^2)
\right]
(2\pi)^n \delta^n(p_1+p_2-K-q_1-\,\dots\, q_m)\nn\;.
\eeeq

In the soft limit we have
\beq
 \delta^n(K-k_1-\,\dots\, k_l)\,\delta^n(p_1+p_2-K-q_1-\,\dots\, q_m)\soft
 \delta^n(p_1+p_2-k_1-\,\dots\, k_l)\,\delta^n(p_1+p_2-K-q_1-\,\dots\, q_m)\;,
\eeq
where the symbol $\soft$ indicates that the equality is valid when the emitted QCD partons are soft. Within this approximation we obtain in the first line of Eq.~(\ref{ps2}) the corresponding leading-order  phase-space $d\text{PS}^{(0)}$, which contains the dependence on the internal variables of the system $F$.
The second line in Eq.~(\ref{ps2}) is the phase-space of a process with one particle of invariant mass $Q$ in the final state plus $m$ soft partons, $d\text{PS}^{2\rightarrow 1+m\,\text{soft}}$.
Then Eq.~(\ref{ps2}) can be rewritten in the following way:
\beq\label{psfact}
\int d\text{PS}\soft\frac{dQ^2}{2\pi}\int d\text{PS}^{(0)} \int d\text{PS}^{2\rightarrow 1+m\,\text{soft}}\;,
\eeq
arriving to a factorized expression for the phase-space in the soft limit.

\section{NLO}
\label{nlo}

At NLO we have to consider the one-loop corrections to the partonic subprocess $a{\bar a} \to F$, and also the real gluon emission subprocess, $a{\bar a} \to F + g$. We begin by computing the latter.

Let $\cm^{(0)}$ be the LO matrix element, 
and $\cm_g^{(0)}$ the correction corresponding to the real gluon emission subprocess at tree-level.
In the limit where the momentum $q$ of the gluon becomes soft, $\cm_g^{(0)}$ can be written in the following factorized way \cite{Bassetto:1984ik}:
\beq\label{gsoftfact}
\big| \cm_g^{(0)}(q,p_1,p_2) \big|^2 \soft
 \left(\ao\,\mu_0^{2\ep}\right)
  8 \pi \, {\cal S}_g(q,p_1,p_2)\,C_a\, 
\;\big| \cm^{(0)}(p_1,p_2) \big|^2 \;,
\eeq 
where $p_1$ and $p_2$ are the momenta of the incoming QCD-partons, and the dependence of the matrix elements on other non-QCD particles momenta is understood.
The symbol $\ao$ stands for the bare coupling constant, and $\mu_0$ is the dimensional-regularization scale.
Renormalization is achieved by the replacement
\beq\label{a0as}
\ao\, \mu_0^{2\epsilon} S_{\epsilon}=\as\,\mu_R^{2\epsilon}
\left(1-\as\,\frac{\beta_0}{\epsilon}+\mathcal{O}(\as^2)\right)\;,
\eeq
where $\beta_0$ is the first coefficient of the QCD beta function and $S_\ep$ is the typical phase-space volume factor in $n=4-2\ep$ dimensions:
\beq
\beta_0=\f{11C_A-2 N_f}{12\pi}\;,\;\;\;\;\;\;\;\;\;\;
S_{\epsilon}=(4\pi)^{\epsilon}\,e^{-\epsilon\gamma_E}\;,
\eeq
being $\gamma_E=0.5772\dots$ the Euler number.

The scalar eikonal function ${\cal S}_g(q,p_1,p_2)$ contains all the dependence of $\cm_g^{(0)}$ on the soft gluon momentum, and takes the form \cite{Bassetto:1984ik}
\beq
{\cal S}_g(q,p_1,p_2)=\f{p_1\cdot p_2}{(p_1\cdot q)(p_2\cdot q)}\;,
\eeq
while the constant $C_a$ depends on the nature of the radiating parton, being
\beq
C_a =
\begin{cases}
 C_A=N_c, & \text{if }a=g \\
 C_F=(N_c^2-1)/(2N_c), & \text{if }a=q,{\bar q}\;,
\end{cases}
\eeq
where $N_c=3$ is the number of colours.

Combining Eq.~(\ref{gsoftfact}) with the phase-space factorization of Eq.~(\ref{psfact}) we arrive at the following expression for the NLO tree-level real gluon emission cross section $\hat{\sigma}_g^{(0)}$:
\beq\label{sigg}
\frac{d\hat{\sigma}_g^{(0)}}{dQ^2}\soft\frac{\hat{\sigma}_0}{2\pi}
 \left(\ao\,\mu_0^{2\ep}\right)
8\pi\, C_a
\int \mathcal{S}_g(q,p_1,p_2)\,
d\text{PS}^{2\rightarrow 1+1\,\text{soft}}\;.
\eeq

The phase-space integration in Eq.~(\ref{sigg}) can be performed in a closed form. After some simple algebra we arrive at
\beq\label{Sgint}
\int {\cal S}_g(q,p_1,p_2)\, d\text{PS}^{2\rightarrow 1+1\,\text{soft}}=
\int \f{p_1\cdot p_2}{(p_1\cdot q)( p_2\cdot q)}\,
\f{\delta^+(K^2-Q^2)}{(2\pi)^{n-2}}\,
\f{ d^{n-1} q}{2\,|q|}\;.
\eeq
Using polar coordinates in $n$ dimensions we can write
\beq
d^{n-1}q=d|q|\,|q|^{n-2}\,d\Omega_{n-2}\;,
\eeq
with
\beq\label{angsolido}
d\Omega_{n-2}\equiv
\sin^{n-3}\theta_1\,\sin^{n-4}\theta_2\dots\sin\theta_{n-3}\,d\theta_1\,d\theta_2\dots d\theta_{n-2}\;,
\eeq
where $\theta_{n-2}\in [0,2\pi)$ and $\theta_i\in [0,\pi]$ for the others.
Parametrizing the momenta in the center-of-mass of the incoming partons, and setting $n=4-2\ep$, it can be shown that
\beq
\int {\cal S}_g(q,p_1,p_2)\, d\text{PS}^{2\rightarrow 1+1\,\text{soft}}=
\f{(16\pi)^\ep}
{s^{1+\ep}(1-z)^{1+2\ep}\,2\pi\,\Gamma(1-\ep)}
\int\f{\sin^{1-2\ep}\theta}
{(1-\cos\theta)(1+\cos\theta)}\,d\theta\;,
\eeq
where $\theta\equiv\theta_1$ is the angle between the soft gluon and the $n$-th axis. Using the variable $y\equiv(1+\cos\theta)/2$, the remaining integral can be carried out as a particular case of
\beq
\int_0^1 y^{\alpha-1}(1-y)^{\beta-1}dy
=
\f{\Gamma(\alpha)\Gamma(\beta)}
{\Gamma(\alpha+\beta)}\;.
\eeq

Finally, the tree-level real gluon emission cross section in the soft limit has the following expression:
\beq\label{gsoft}
\frac{d\hat{\sigma}_g^{(0)}}{dQ^2}\soft\
-\frac{\hat{\sigma}_0}{s}
 \left(\f{\ao}{2\pi}\right)
\frac{\left(4\pi\mu_0^2/s\right)^{\epsilon}}{(1-z)^{1+2\epsilon}}\,\,
C_a\,
 \frac{4\ \Gamma(1-\epsilon)}{\epsilon\ \Gamma(1-2\epsilon)}\;.
\eeq
This formula is valid to all orders in $\ep$ for any reaction of the kind of Eq.~(\ref{proceso}), and its only dependence on a particular process is in the Born-level cross section $\hat{\sigma}_0$. The $C_a$ factor only depends on the nature of the incoming partons. The expansion of the factor $(1-z)^{-1-2\epsilon}$ leads to the appearance of $\delta(1-z)$ and ${\cal D}_i(z)$ terms, according to the following relation:
\beq
\f{1}{(1-z)^{1+a\ep}}=-\f{1}{a\ep}\,\delta(1-z)+\sum_{i=0}^\infty \f{(-a\ep)^i}{i!}\,{\cal D}_i(z)\;.
\eeq

We now have to evaluate the one-loop correction term. Even though this cannot be done in a process independent way, the infrared-singular behaviour of QCD amplitudes at one-loop (and two-loop) order is well known \cite{1L sing1,1L sing2,1L sing3,Catani:1998bh}. For the processes we are interested in, the renormalized one-loop order amplitude $\cm^{(1)}$ can be written in terms of the Born-level amplitude $\cm^{(0)}$ in the following way\footnote{For the sake of simplicity we omit the explicit dependence of the matrix elements on the partons momenta.} \cite{Catani:1998bh}:
\beq\label{M1l}
\cm^{(1)}(\ep) = \frac{\as}{2\pi}\left[
\bom{I}_a^{(1)}(\ep)
\cm^{(0)}(\ep)
+  \cm^{(1)}_{\rm fin}(\ep)\,\right]\;,
\eeq
where
\beq\label{I1}
\bom{I}_a^{(1)}(\ep)=
-\left(-\f{4\pi\mu_R^2}{s}\right)^\ep
\frac{S_\ep^{-1}}{\Gamma(1-\ep)} 
\left(
 C_a\, \frac{1}{\ep^2} 
+ \gamma_a \;\frac{1}{\ep}
\right)
\;,
\eeq
and the contribution $\cm^{(1)}_{\rm fin}$ is finite when $\ep\to 0$ \footnote{We explicitly keep higher order terms in $\ep$ originated in the one-loop amplitude as they contribute to the final result to the same order in the dimensional regularization parameter. Those in $\cm^{(0)}$ are implicitly included in the definition of $\hat{\sigma}_0=\f{1}{2s}\int \big|\cm^{(0)}\big|^2 d\text{PS}^{(0)}$.}. The coefficient $\gamma_a$ depends on the initial-state partons, being
\beq
\gamma_a=
\begin{cases}
 \tfrac{11}{6}C_A-\tfrac{1}{3}N_f, & \text{if }a=g \\
 \;\tfrac{3}{2} C_F, & \text{if }a=q,{\bar q}\;.
\end{cases}
\eeq

For the NLO calculation we only need the ${\cal O}(\as)$ term of the squared matrix element, that is $\cm^{(1)}(\cm^{(0)})^*+(\cm^{(1)})^*\cm^{(0)}$. Performing the formal phase-space integration of this term we arrive at the following expression for the one-loop virtual contribution to the cross section:
\beeq
\frac{d\hat{\sigma}^{(1)}_v}{dQ^2}\!\!\!&=&\!\!\!\frac{\hat{\sigma}_0}{s}\label{virtual1}
\left(\f{\as}{2\pi}\right)
\delta(1-z)\,\bigg\{\!\!
-\!\left(\f{4\pi\mu_R^2}{s}
\right)^\ep
\f{S_\ep^{-1}\,\Gamma(\ep)}{\Gamma(1-2\ep)\Gamma(2\ep)}
\left( C_a\, \frac{1}{\ep^2} 
+ \gamma_a \;\frac{1}{\ep}\right)
+\f{\hat{\sigma}^{(1)}_{\text{fin}}(\ep)}{\hat{\sigma}_0}\bigg\}\;,
\eeeq
where $\hat{\sigma}^{(1)}_{\text{fin}}(\ep)$ is a one-loop finite contribution to the cross section defined by 
\beq
\f{\hat{\sigma}^{(1)}_{\text{fin}}(\ep)}{\hat{\sigma}_0}=
\int
\left(\cm^{(1)}_\text{fin}\left(\cm^{(0)}\right)^*+
\left(\cm^{(1)}_\text{fin}\right)^*\cm^{(0)}
\right)
d\text{PS}^{(0)}
\bigg/ \int\big|\cm^{(0)}\big|^2
d\text{PS}^{(0)}\;.\label{sigfin1}
\eeq

To obtain the NLO soft-virtual contribution to the coefficient function $G_{a\bar{a}}(z)$ we still have to add to Eqs.~(\ref{gsoft}) and (\ref{virtual1}) the counterterms coming from mass factorization.
Keeping terms up to second order in powers of $\as$, relevant for the NNLO calculation in the next section, we have
\beeq
\f{d\hat{\sigma}(s)}{dQ^2}\!\!&=&\!\!\f{\hat{\sigma}_0}{s}\;\Bigg\{
\delta(1-x)
+\f{\as}{2\pi}\left\{
G_{a\bar{a}}^{(1)}(z)-\f{2}{\ep}P_{aa}^{(0)}(z)\right\}\nn\\
&&\!\!\!\!\!\!+\,\left(\f{\as}{2\pi}\right)^2
\bigg\{\bigg[
2\left(P_{aa}^{(0)}\!\otimes P_{aa}^{(0)}\right)(z)
+2\pi\beta_0 P_{aa}^{(0)}(z)\bigg]\f{1}{\ep^2}\label{renormalizado}\\
&&\;\;\;\;\;\;\;\;\;\;\;-\,\bigg[P_{aa}^{(1)}(z)+2\left(P_{aa}^{(0)}\!\otimes G_{a\bar{a}}^{(1)}\right)(z)
\bigg]\f{1}{\ep}+G^{(2)}_{a\bar{a}}(z)\nn
\bigg\}+{\cal O}(\as^3)\;
\Bigg\},
\eeeq
where for simplicity we have set $\mu_F=Q$. The symbol ``$\otimes$" stands for the usual convolution.
The Altarelli-Parisi splitting functions in the soft limit ($z\to 1$) take the form \cite{Ellis:1991qj}
\beeq
P_{aa}^{(0)}(z)&\soft& \gamma_a\, \delta(1-z)+2\, C_a\, {\cal D}_0(z)\;,\\
P_{aa}^{(1)}(z)&\soft& \gamma^{(1)}_a\, \delta(1-z)+C_a
\left[ C_A \left(\f{67}{9}-\f{\pi^2}{3}\right)
- N_f\; \f{10}{9}
\right] {\cal D}_0(z)\;,
\eeeq
where we have defined
\beq
\gamma_a^{(1)}=
\begin{cases}
 C_A^2 \left(\f{8}{3}+3\zeta_3\right)-\f{1}{2} C_F N_f-\f{2}{3} C_A N_f, & \text{if }a=g \vspace*{0.3cm}\\
 C_F^2\left(\f{3}{8}-\f{\pi^2}{2}+6\zeta_3\right)\!
+C_F C_A \left(\f{17}{24}+\f{11\pi^2}{18}-3\zeta_3\right)\!
-C_F N_f \left(\f{1}{12}+\f{\pi^2}{9}\right), & \text{if }a=q,{\bar q}\;.
\end{cases}
\eeq

Combining the results of Eqs.~(\ref{gsoft}) and (\ref{virtual1}) with the counterterms coming from Eq.~(\ref{renormalizado}) we arrive at a closed expression valid to all orders in $\ep$ for the NLO soft-virtual coefficient function $G_{a\bar{a}}^{(1)}$. For simplicity, we only write the first three terms of its expansion in powers of $\ep$ \footnote{Contributions up to ${\cal O}(\ep^2)$ are needed to build the renormalization and factorization counterterms for a calculation to N$^3$LO accuracy.}:
\beeq
G^{(1)}_{a\bar{a}}(z)&\soft&\delta(1-z)\;\Bigg\{
C_a\,\f{2\pi^2}{3}
+\frac{\hat{\sigma}^{(1)}_{\text{fin}(0)}}{\hat{\sigma}_0}
+\bigg[\!
-4C_a\,\zeta_3+\gamma_a\f{7\pi^2}{6}+\f{\hat{\sigma}^{(1)}_{\text{fin}(1)}}{\hat{\sigma}_0}\,
\bigg]\ep\nn\\
&&\;\;\;\;\;\;\;\;\;\;\;\;\;\;\;\;\;\;\;\;\;\;\;\;\;\;\;
+\;\bigg[
-C_a\f{17\pi^4}{90}+\gamma_a\f{2\zeta_3}{3}+
\f{\hat{\sigma}^{(1)}_{\text{fin}(2)}}{\hat{\sigma}_0}\,
\bigg]\ep^2
\Bigg\}\\
&&\!\!\!\!\!\!\!\!\!\!\!\!\!\!\!\!\!\!\!\!
+\;8\,C_a\,{\cal D}_1(z)+
C_a\bigg(
\pi^2{\cal D}_0(z)-8\,{\cal D}_2(z)
\bigg)\ep
+C_a\bigg(
\f{28\zeta_3}{3}{\cal D}_0(z)-2\pi^2{\cal D}_1(z)+\f{16}{3}{\cal D}_3(z)
\bigg)\ep^2
+{\cal O}(\ep^3)\;,\label{G1}\nn
\eeeq
where we have set $\mu_R=\mu_F=Q\soft \sqrt{s}$.
The dependence of this expression on a particular process is contained only
in the one-loop coefficients $\hat{\sigma}^{(1)}_{\text{fin}(i)}$ defined by
\beq
\hat{\sigma}^{(1)}_{\text{fin}}(\ep)=\sum_{i=0}^\infty \hat{\sigma}^{(1)}_{\text{fin}(i)}\,\ep^i\;.
\eeq

\section{NNLO}\label{nnlo}

At second order in the perturbative expansion we have to consider the four ${\cal O}(\as^2)$ parton subprocesses of Eq.~(\ref{procesos}).
The one-loop correction to the subprocess $a{\bar a} \to F+g$ can be obtained in a very similar way to the tree-level one. Let $\cm_g^{(1)}$ be the one-loop gluon emission matrix element. The analogous to Eq.~(\ref{gsoftfact}) is given by the soft limit of one-loop amplitudes as \cite{Catani:2000pi,Bern:1995ix,Bern:1998sc,Bern:1999ry,Catani:2004nc}
\beeq
&&\!\!\!\!\! \!\!\!\!\!\!\!\!
\left(\cm^{(1)}_g\right)^* \, \cm^{(0)}_g +
  \cm^{(1)}_g \, \left(\cm^{(0)}_g\right)^*
\soft
 \left( \ao \mu_0^{2\epsilon} \right) C_a \nn\,
\label{1loopsquare2}
\bigg\{ 8\pi\,{\cal S}_g(q)\Big[ 
\left(\cm^{(1)}\right)^* \, \cm^{(0)} +
  \cm^{(1)} \, \left(\cm^{(0)}\right)^* 
 \Big]\\
&& \;\;\;\;\;\;\;\;\;\;\;\;\;\;\;\;\;
 -\left( \ao \mu_0^{2\epsilon} \right) C_A\; \frac{(2\pi)^\ep}{\ep^2} \;\f{2\,\Gamma^4(1-\ep)
\Gamma^2(1+\ep) \Gamma(\ep)}{\Gamma^2(1-2\ep)\Gamma(2\ep)}
\, \big({\cal S}_g(q)\big)^{1+\ep}
\big|\cm^{(0)}\big|^2 \bigg\}\;.
\eeeq
The phase-space integrals we have to perform are
\beq
\int \mathcal{S}_g(q,p_1,p_2)\,
d\text{PS}^{2\rightarrow 1+1\,\text{soft}}\;
\;\;\;\;\;\;\;\;\text{and}\;\;\;\;\;\;\;\;
\int \left[\mathcal{S}_g(q,p_1,p_2)\right]^{1+\ep}\,
d\text{PS}^{2\rightarrow 1+1\,\text{soft}}\;,
\eeq
and their calculation can be achieved with the tools discussed in section \ref{nlo}. The result is
\beeq\label{sigmag(1)}
\frac{d \hat{\sigma}^{(1)}_g}{dQ^2}\!&\soft&\!
-\frac{\hat{\sigma}^{(1)}_v}{s}
\left(\frac{\ao}{2\pi}\right)
\frac{\left(4\pi \mu_0^2/s\right)^{\epsilon}}{(1-z)^{1+2\epsilon}}\,\,
C_a\,  \frac{4\ \Gamma(1-\epsilon)}{\epsilon\ \Gamma(1-2\epsilon)}\\
&-&\!\!\frac{\hat{\sigma}_0}{s}
\left(\frac{\ao}{2\pi}\right)^2
\frac{\left(4\pi \mu_0^2/s\right)^{2\epsilon}}{(1-z)^{1+4\epsilon}}\,\,
C_a\,C_A\,
\f{2\, \Gamma^3(1-\ep)\Gamma^3(\ep)\Gamma(-2\ep)}
{\Gamma(1-4\ep)\Gamma(1-2\ep)\Gamma(2\ep)}\;,\nn
\eeeq
where $\hat{\sigma}^{(1)}_v$ can be written as in Eq.~(\ref{virtual1}), using the delta function to perform the $Q^2$ integral.

We continue by computing the double real emission subprocesses, that is $a\bar{a}\to F+q+{\bar q}$ and $a\bar{a}\to F+g+g$. For the NNLO squared amplitudes the following infrared factorization formulae hold \cite{Catani:1999ss,Berends:1988zn}:
\beeq
\label{qqsoftfac2}
\big| \cm_{q{\bar q}}(q_1,q_2,p_1,p_2) \big|^2 \!\!\! &\soft & \!\!\!
 \big(\ao \mu_0^{2\ep}\big)^2\, 8\pi^2
\,C_a
\big[\,
{\cal I}_{11}(q_1,q_2)+{\cal I}_{22}(q_1,q_2)-2{\cal I}_{12}(q_1,q_2)\,
\big]
\big| \cm_0(p_1,p_2) \big|^2 ,\\ \nn \\
\label{Mggsoft}
\big| {\cal M}_{gg}(q_1,q_2,p_1,p_2) \big|^2 \!\!\! &\soft & \!\!\!
\big(\ao \mu_0^{2\ep} \big)^2\,
16\pi^2
  \big| {\cal M}_0(p_1,p_2) \big|^2 
\\&& \!\!\!\! \times 
\Big[ 4\, C_a^2 \, {\cal S}_g(q_1) \;
{\cal S}_g(q_2) \;
- \,C_A\, C_a
\Big( {\cal S}_{11}(q_1,q_2)+{\cal S}_{22}(q_1,q_2)-2\,{\cal S}_{12}(q_1,q_2)
\Big)
\Big] \nn \;,
\eeeq
where all the dependence on the momenta $q_1$ and $q_2$ of the soft particles is embodied in the functions ${\cal I}_{ij}$ and ${\cal S}_{ij}$, which take the form \cite{Catani:1999ss}
\beeq
{\cal I}_{ij}(q_1,q_2) &=&
- \;\frac{2 (p_i \cdot p_j) \,(q_1 \cdot q_2)
+ [p_i \cdot (q_1 - q_2)]\, [p_j \cdot (q_1 - q_2)]}{2 (q_1 \cdot q_2)^2 
\,[p_i\cdot (q_1+q_2)]\, [p_j \cdot (q_1+q_2)]}\;,\\ \nn\\
{\cal S}_{ij}(q_1,q_2) &=& \f{(1-\ep)}{(q_1 \cdot q_2 )^2} \;
\f{p_i \cdot q_1 \,p_j \cdot \,q_2 + p_i \cdot q_2 \,p_j \cdot \,q_1}
{p_i\cdot (q_1+q_2) \; p_j\cdot (q_1+q_2)} \nn \\
\label{dsoftfun}
&-& \f{(p_i \cdot p_j)^2}{2 p_i\cdot q_1 \; p_j\cdot q_2 \;
p_i\cdot q_2 \; p_j\cdot q_1}
\left[ 2 - \f{p_i \cdot q_1 \,p_j \cdot \,q_2 + p_i \cdot q_2 \,p_j \cdot \,q_1}
{p_i\cdot (q_1+q_2) \; p_j\cdot (q_1+q_2)} \right] \\
&+& \f{p_i\cdot p_j}{2 q_1 \cdot q_2}
\left[ \f{2}{p_i\cdot q_1 \,p_j \cdot \,q_2} + 
       \f{2}{p_j\cdot q_1 \,p_i \cdot \,q_2} \right. \nn \\
&-& \left.
       \f{1}{p_i\cdot (q_1+q_2) \; p_j\cdot (q_1+q_2)}
   \left( 4 + 
  \f{(p_i \cdot q_1 \,p_j \cdot \,q_2 + p_i \cdot q_2 
  \,p_j \cdot \,q_1)^2}{p_i\cdot q_1 \; p_j\cdot q_2 \;
  p_i\cdot q_2 \; p_j\cdot q_1}
\right) \right] \;.
\eeeq

To perform the phase-space integration of both contributions we use the parametrization of $d\text{PS}^{2\rightarrow 1+2\,\text{soft}}$ introduced in \cite{Matsuura:1988sm} for the calculation of the second order corrections to the Drell-Yan process. Introducing the variables $q=q_1+q_2$ and $s_{12}=q^2$ we can write
\beeq
\int d\text{PS}^{2\rightarrow 1+2\,\text{soft}}&=&\f{1}{(2\pi)^{2n-3}}\nn\\
&\times&\int d^n K \int d^n q\int ds_{12}
\delta^+(K^2-Q^2)\delta^+(q^2-s_{12})\delta^n(p_1+p_2-K-q)\nn\\
&\times&\int d^n q_1 \int d^n q_2 \delta^+(q_1^2) \delta^+(q_2^2) \delta^n(q-q_1-q_2)\;.
\eeeq
The last line of the above equation is most easily computed in the center-of-mass of $q_1$ and $q_2$. In this frame, orientated so that $p_1$ is in the direction of the $n\text{-th}$ axis and $p_2$ lies in the plane defined by the $n\text{-th}$ and $(n-1)\text{-th}$ axes, the momenta can be parametrized as follows \cite{Matsuura:1988sm}:
\beeq
q_1&=&\tfrac{1}{2}\sqrt{s_{12}}\,(1,\dots,\cos\phi\sin\theta,\cos\theta)\;,\nn\\
q_2&=&\tfrac{1}{2}\sqrt{s_{12}}\,(1,\dots,-\cos\phi\sin\theta,-\cos\theta)\;,\nn\\
p_1&=&\frac{(s-\tilde{t}\,)}{2\sqrt{s_{12}}}\,(1,0,\dots,0,0,1)\;,\nn\\
K&=&\left(\f{s-Q^2-s_{12}}{2\sqrt{s_{12}}},0,\dots,0,
|K|\sin\psi,|K|\cos\psi\right)\;,\nn\\
|K|&=&\f{\sqrt{\lambda(s,Q^2,s_{12})}}{2\sqrt{s_{12}}}\nn\;,\\
\cos\psi&=&\f{(s-Q^2)(\tilde{u}-Q^2)-s_{12}(\tilde{t}+Q^2)}
{(s-\tilde{t}\,)\sqrt{\lambda(s,Q^2,s_{12})}}\;,\label{param2}
\eeeq
where we have defined $\theta\equiv \theta_1$ and $\phi\equiv \theta_2$ (see Eq.~(\ref{angsolido})), and
\beq\label{defvn}
\tilde{t}=2\,p_1\cdot K,\;\;\;\;\; \tilde{u}=2\,p_2\cdot K, \;\;\;\;\;
s=(p_1+p_2)^2, \;\;\;\;\; s_{12}=s-\tilde{t}-\tilde{u}+Q^2,
\eeq
where  $\lambda(a,b,c)$ is the K\"{a}llen function, $\lambda(a,b,c)=a^2+b^2+c^2-2ab-2bc-2ca$.
The dots in $q_1$ and $q_2$ represent $n-3$ unspecified components of momentum, which are trivially integrated with the present parametrization.
Using momentum conservation, we obtain for $p_2$ the following expression:
\beeq
\label{paramp2}
p_2&=&\f{s_{12}+\tilde{t}-Q^2}{2\sqrt{s_{12}}}\;
(1,0,\dots,0,\sin\chi,\cos\chi)\;,\nn\\
\cos\chi&=&
\f{2\sqrt{s_{12}}}{s_{12}+\tilde{t}-Q^2}
\left(
|K|\cos\psi+\f{\tilde{t}-s}{2\sqrt{s_{12}}}
\right)\;.
\eeeq

Considering the above parametrization and introducing the variables $x$, $y$ and $z$:
\beeq
z&=&Q^2/s\;,\nn\\
\tilde{u}&=&s(1-y(1-z))\;,\nn\\
\tilde{t}&=&s\left(
z+y(1-z)-\f{y(1-y)x(1-z)^2}{1-y(1-z)}
\right)\;,
\eeeq
the double real emission phase-space can be written as follows \cite{Matsuura:1988sm}:
\beeq\label{psexacto}
\int  d\text{PS}^{2\rightarrow 1+2\,\text{soft}}\!\!\!&=&\!\!\!\nn
\f{1}{(4\pi)^n}\,\f{s^{n-3}}{\Gamma(n-3)}\,
(1-z)^{2n-5}
\int_0^\pi d\theta \int_0^\pi d\phi\,
\sin^{n-3}\theta\; \sin^{n-4}\phi\\
&\times&\!\!\!\!\!\!\int_0^1 dy \int_0^1 dx\,
[y(1-y)]^{n-3}\,
[x(1-x)]^{n/2-2}\,
[1-y(1-z)]^{1-n/2}\,.
\eeeq

Up to this point we have kept the exact expression for the phase-space. In the soft limit the last factor of Eq.~(\ref{psexacto}) can be approximated by $1$, obtaining
\beeq
\int  d\text{PS}^{2\rightarrow 1+2\,\text{soft}}\!\!\!&\soft &\!\!\!\nn
\f{1}{(4\pi)^n}\,\f{s^{n-3}}{\Gamma(n-3)}\,
(1-z)^{2n-5}
\int_0^\pi d\theta \int_0^\pi d\phi\,
\sin^{n-3}\theta\; \sin^{n-4}\phi\\
&\times&\int_0^1 dy \int_0^1 dx\,
[y(1-y)]^{n-3}\,
[x(1-x)]^{n/2-2}\,.
\eeeq
With this approximation the integrals we have to perform become considerably simpler.

We now have to express the integrands in terms of the variables $x$, $y$, $z$, $\theta$ and $\phi$. The quark-antiquark emission subprocess is relatively simple. The ${\cal I}_{11}(q_1,q_2)$ term of Eq.~(\ref{qqsoftfac2}), for example, takes the following form
\beq
{\cal I}_{11}(q_1,q_2)=-
\f{2\,\cos^2\theta}{(s_{12})^2}=-
\f{2\,(1-(1-z)\,y)^2\cos^2\theta}
{s^2 (1-z)^4 \,y^2(1-y)^2\, x^2}\soft -
\f{2\,\cos^2\theta}
{s^2 (1-z)^4 \,y^2(1-y)^2\, x^2}\;,
\eeq
where we have considered the $z\to 1$ limit in the numerator in the last step. Then the phase-space integral can be done in a direct way, obtaining
\beq\label{I11int}
\int {\cal I}_{11}(q_1,q_2)\, d\text{PS}^{2\rightarrow 1+2\,\text{soft}}\soft
-\f{(4\pi)^{2\ep}}{s^{1+2\ep}\, (1-z)^{1+4\ep}}\,
\f{\Gamma(-1-\ep)\,\Gamma(-2\ep)\,\Gamma(2-\ep)}
{32\pi^3\,\Gamma(-4\ep)\,\Gamma(4-2\ep)}\;.
\eeq
The phase-space integral of the ${\cal I}_{22}(q_1,q_2)$ term yields the same result, since the integrand can be obtained from ${\cal I}_{11}$ by the exchange $p_1\leftrightarrow p_2$. The same happens for two integrands that only differ in the exchange of $q_1\leftrightarrow q_2$.

The remaining term is ${\cal I}_{12}(q_1,q_2)$. It can be split into two contributions, ${\cal I}_{12}={\cal I}_{A}+{\cal I}_{B}$, where
\beeq
{\cal I}_{A}(q_1,q_2)\!\!\!\!&=&\!\!\!\!
 \frac{- (p_1 \cdot p_2) \;(q_1 \cdot q_2)}
{ (q_1 \cdot q_2)^2 
\,[p_1\cdot (q_1+q_2)]\, [p_2 \cdot (q_1+q_2)]}\soft
-\f{4}
{s^2\,(1-z)^4\,y^2(1-y)^2\,x}\;,\\ \nn \\
{\cal I}_{B}(q_1,q_2)\!\!\!\!&=&\!\!\!\!
\frac{-[p_1 \cdot (q_1 - q_2)]\, [p_2 \cdot (q_1 - q_2)]}{2 (q_1 \cdot q_2)^2 
[p_1\cdot (q_1+q_2)] [p_2 \cdot (q_1+q_2)]}\soft
\f{2x\cos\phi\,\sin 2\theta-(2-4x)\cos^2\theta}
{s^2\,(1-z)^4\,y^2(1-y)^2\,x^2}\;.
\eeeq
Their integration is straightforward and the result is
\beeq
\int{\cal I}_A(q_1,q_2)\,\label{Iaint}
 d\text{PS}^{2\rightarrow 1+2\,\text{soft}}&\soft&
-\f{(4\ep)^{2\ep}}{s^{1+2\ep}\, (1-z)^{1+4\ep}}\,
\f{\Gamma(-\ep)\,\Gamma(-2\ep)\,\Gamma(1-\ep)}{16\pi^3\,\Gamma(1-4\ep)\,\Gamma(2-2\ep)}\;,\\
\int{\cal I}_B(q_1,q_2)\,
 d\text{PS}^{2\rightarrow 1+2\,\text{soft}}&\soft&\label{Ibint}
-\f{(4\pi)^{2\ep}}{s^{1+2\ep}\, (1-z)^{1+4\ep}}\,
\f{\Gamma(-1-\ep)\,\Gamma(-2\ep)\,\Gamma(2-\ep)}
{8\pi^3\,\Gamma(1-4\ep)\,\Gamma(4-2\ep)}\;.
\eeeq
Combining Eqs.~(\ref{I11int}), (\ref{Iaint}) and (\ref{Ibint}) with Eq.~(\ref{qqsoftfac2}) we arrive at a rather compact expression for the NNLO double real soft quark-antiquark emission cross section:
\beq\label{sigmaqq}
\frac{d \hat{\sigma}_{q\bar{q}}}{dQ^2}\soft\frac{\hat{\sigma}_0}{s}\left(\frac{\ao}{2\pi}\right)^2
\frac{\left(4\pi\mu_0^2/s\right)^{2\epsilon}}{(1-z)^{1+4\epsilon}}\,\,
C_a\,N_f\,
\f{2\,\Gamma^2(2-\ep)\,\Gamma(-2\ep)}{\ep^2\,\Gamma(4-2\ep)\,\Gamma(-4\ep)}\;.
\eeq
The $N_f$ factor arises from the sum over the different $q{\bar q}$ pair flavours.

\vspace{0.5cm}

The integration of the double gluon emission matrix element of Eq.~(\ref{Mggsoft}) is more cumbersome. 
The following formula is used several times in the calculation of the angular integrals \cite{Matsuura:1988sm}:
\beeq
\int_0^\pi d\!\!\!\!\!\!&\theta&\!\!\!\!\! \int_0^\pi d\phi\; \frac{\sin^{n-3}\theta\,\sin^{n-4}\phi}{(1-\cos\theta)^i(1-\cos\xi\cos\theta-\sin\xi\cos\phi\sin\theta)^j}\label{integralvn}\\
&=&\!2^{1-i-j}\pi\,\frac{\Gamma(n/2-1-j)\Gamma(n/2-1-i)}{\Gamma(n-2-i-j)}\,
\frac{\Gamma(n-3)}{\Gamma^2(n/2-1)}\,
F\left(i,j;{n\over 2}-1;\cos^2{\xi \over 2}\right)\nn\,,
\eeeq
where $F(a,b;c;z)$ is the hypergeometric function.
 
As an example we show in detail the integration of the simplest term, $\int  {\cal S}_g(q_1) \,{\cal S}_g(q_2) d\text{PS}^{2\rightarrow 1+2\,\text{soft}}$.
After partial fractioning, it can be written in the following way
\beq\label{SgSg2}
{\cal S}_g(q_1) \,{\cal S}_g(q_2)=
\f{s^2}{(s-\tilde{t}\,)\,(s_{12}+\tilde{t}-Q^2)}
\left(\,
\f{1}{p_1\cdot q_1}+\f{1}{p_1\cdot q_2}\,
\right)
\left(\,
\f{1}{p_2\cdot q_1}+\f{1}{p_2\cdot q_2}\,
\right)\,.
\eeq 
Out of the four terms we obtain multiplying the factors in parenthesis we have to integrate only two because the others can be obtained from them by the exchange $p_1\leftrightarrow p_2$ or $q_1\leftrightarrow q_2$.

Let ${\cal {S}}_A(q_1,q_2)$ be the term corresponding to the product $(p_1\cdot q_1)(p_2\cdot q_1)$ in the denominator. Using the parametrization of Eq.~(\ref{param2}) we arrive at
\beq
{\cal {S}}_A(q_1,q_2)=\left(
\f{4s}{(s-\tilde{t}\,)\,(s_{12}+\tilde{t}-Q^2)}
\right)^2\left(
\f{1}
{(1-\cos\theta)
(1-\cos\chi\cos\theta-\sin\chi\cos\phi\sin\theta)}\right)\,.
\eeq
The first factor can be written in terms of the variables $x$, $y$ and $z$. Then the second factor can be integrated using Eq.~(\ref{integralvn}) with $i=j=1$ and $\xi=\chi$ (the factors $\sin^{n-3}\theta\,\sin^{n-4}\phi$ are present in $ d\text{PS}^{2\rightarrow 1+2\,\text{soft}}$). The argument of the hypergeometric function, $\cos^2\f{\chi}{2}$, can be written using the definition of $\cos\chi$ in Eq.~(\ref{paramp2}) in the following way:
\beq
\cos^2\f{\chi}{2}=
\f{1+\cos\chi}{2}\soft
1+x\;,
\eeq
where the last step is valid in the soft limit, since $\cos\chi\to 1-2x$ when $z\to 1$.

Now we have to perform the integration in the variables $x$ and $y$. The first one takes the following form in the soft limit
\beq\label{intz}
\int_0^1
F(1,1;\tfrac{n}{2}-1;1-x)
[x(1-x)]^{n/2-2}dx\; ,
\eeq
which is a particular case of
\beq
\int_0^1F(1,1,\gamma,t)[(1-t)t]^{\gamma-1}dt=
\f{\Gamma^2(\gamma)\Gamma(2\gamma-2)}{\Gamma^2(2\gamma-1)}\;.
\eeq
The remaining integral can be done straightforwardly, and the final result is
\beq\label{saint}
\int {\cal S}_A(q_1,q_2)\,
 d\text{PS}^{2\rightarrow 1+2\,\text{soft}}\soft
-\f{(4\pi)^{2\ep}}
{s^{1+2\ep}(1-z)^{1+4\ep}}\,
\f{\Gamma^2(-\ep)}{32\pi^3\,\ep\,\Gamma(1-4\ep)}\;.
\eeq

Now let ${\cal S}_B(q_1,q_2)$ be the term of Eq.~(\ref{SgSg2}) corresponding to the product $(p_1\cdot q_1)(p_2\cdot q_2)$ in the denominator. In this case we have
\beq
{\cal {S}}_B(q_1,q_2)=\left(
\f{4s}{(s-\tilde{t}\,)\,(s_{12}+\tilde{t}-Q^2)}\right)^2
\left(\f{1}
{(1-\cos\theta)
(1+\cos\chi\cos\theta+\sin\chi\cos\phi\sin\theta)}\right)\,.
\eeq
To perform the angular integral we use again Eq.~(\ref{integralvn}), taking in this case $i=j=1$ and $\xi=\chi+\pi$. Then the result only differs from the previous one in the argument of the hypergeometric function, which is
\beq
\cos^2\f{\xi}{2}=\cos^2\left(\f{\chi+\pi}{2}\right)=
\f{1-\cos\chi}{2}\soft x\;,
\eeq
and the $x$ integral takes the form
\beq
\int_0^1
F(1,1;\tfrac{n}{2}-1;x)
[x(1-x)]^{n/2-2}dx\;,
\eeq
which is the same of Eq.~(\ref{intz}) through the change of variables $x\to 1-x$. Then the phase-space integral of $\int {\cal S}_B(q_1,q_2)\,d\text{PS}^{2\rightarrow 1+2\,\text{soft}}$ coincides with Eq.~(\ref{saint}), and the four terms of Eq.~(\ref{SgSg2}) give equal contributions to the cross section, yielding
\beq\label{intgg1}
\int {\cal S}_g(q_1)\,{\cal S}_g(q_2)\,
 d\text{PS}^{2\rightarrow 1+2\,\text{soft}}\soft
-\f{(4\pi)^{2\ep}}
{s^{1+2\ep}(1-z)^{1+4\ep}}\,
\f{\Gamma^2(-\ep)}{8\pi^3\,\ep\,\Gamma(1-4\ep)}\;.
\eeq

The integration of the remaining terms of Eq.~(\ref{Mggsoft}) is more complicated due to the more involved expression of the two-gluon eikonal function ${\cal S}_{ij}(q_1,q_2)$. However, the techniques needed for the calculation are essentially the ones we have described so far. The result we obtain is the following:
\beeq\label{intgg2}
\int {\cal S}_{11}(q_1,q_2)+{\cal S}_{22}(q_1,q_2)-2{\cal S}_{12}(q_1,q_2)\,
 d\text{PS}^{2\rightarrow 1+2\,\text{soft}}
 \!\!\!\!&\soft &\!\!\!\!
 -\f{(4\pi)^{2\ep}}{s^{1+2\ep}(1-z)^{1+4\ep}}
 \f{\Gamma(-\ep)\Gamma(-2\ep)\Gamma(2-\ep)}{16\pi^3\ep^2\Gamma(-4\ep)\Gamma(4-2\ep)}
 \nn\\
&&\!\!\!\!\!\!\!\!\!\!\!\!\!\!\!\!\!\!\!\!\!\!\!\!\!\!\!\!\!\!\!\!\!\!\!\!\!\!\!\!\!\!\!\!\!\!\!\!\!\!\!\!\!\!\!\!\!\!\!\!\!\!\!\!\!\!\!\!\!\!
\!\!\!\!\!\!\!\!\!\!\!\!\!\!\!\!\!\!\!\!\!\!\!\!\!\!\!\!
\times\;\Big[-3+(19-11\epsilon)\epsilon+2(3+4\epsilon(\epsilon-2))\,
_3F_2(1,1,-\epsilon;1-2\epsilon,1-\epsilon;1)\Big]\,,
\eeeq
where the generalized hypergeometric function $_3F_2$ arises from the following integral:
\beq
\int_0^1 (1-x)^{-\ep} x^{-1-\ep} F(1,1;1-\ep,x)dx=
\f{\Gamma^2(-\ep)}{2\,\Gamma(-2\ep)}
\, _3F_2(1,1,-\epsilon;1-2\epsilon,1-\epsilon;1)\;.
\eeq

Combining the results of Eqs.~(\ref{intgg1}) and (\ref{intgg2}), adding the factors coming from Eq.~(\ref{Mggsoft}) and a $\tfrac{1}{2}$ factor for identical particles in the final state, we finally arrive at the double soft gluon emission cross section:
\beeq\label{sigmagg}
\frac{d \hat{\sigma}_{gg}}{dQ^2}\!\!&\soft&\!\!\frac{\hat{\sigma}_0}{s}\left(\frac{\ao}{2\pi}\right)^2
\frac{\left(4\pi\mu_0^2/s\right)^{2\epsilon}}{(1-z)^{1+4\epsilon}}\,
\f{\Gamma(-\ep)}{\ep^2\Gamma(-4\ep)}
\,
\bigg\{
2\,C_a^2\, \Gamma(-\ep)
+ C_a\, C_A \,
\f{\Gamma(-2\ep)\Gamma(2-\ep)}{\Gamma(4-2\ep)}\\
&\times& \Big[-3+(19-11\epsilon)\epsilon+2(3+4\epsilon(\epsilon-2))\,
_3F_2(1,1,-\epsilon;1-2\epsilon,1-\epsilon;1)\Big]
\bigg\}\nn\;.
\eeeq
Again, this result is valid at all orders in $\ep$ \footnote{The contribution from the generalized hypergeometric function $_3F_2$ is the only term that can not be written in terms of simpler $\Gamma$ functions \cite{Anastasiou:2012kq}.}. 

\vspace{0.5cm}

We now have to evaluate the ${\cal O}(\as^2)$ virtual corrections. One contribution comes from the square of the matrix element $\cm^{(1)}$ of Eq.~(\ref{M1l}). The other arises from the two-loop (renormalized) matrix element $\cm^{(2)}$, whose infrared-singular behaviour can be written as \cite{Catani:1998bh,Aybat:2006mz}:
\beq\label{M2l}
\cm^{(2)}(\ep) = \frac{\as}{2\pi}
\bom{I}_a^{(1)}(\ep) \;  \cm^{(1)}(\ep) 
+ \left(\frac{\as}{2\pi}\right)^2\Big[ 
\bom{I}_a^{(2)}(\ep)\, \cm^{(0)}(\ep)
+ \cm^{(2)}_{\rm fin}(\ep)\,\Big]\;,
\eeq
where $\cm^{(2)}_{\rm fin}$ is finite when $\ep \to 0$, and the function $\bom{I}_a^{(2)}(\ep)$ is explicitly given by \cite{Catani:1998bh}:
\beeq
\bom{I}_a^{(2)}(\ep)\!\!\!\!&=&\!\!\!\!
\f{\left(-\f{4\pi\mu_R^2}{s}\right)^\ep S_\ep^{-1}}{72\ep^4\Gamma(1-\ep)}
\bigg\{
12\ep(C_a+\ep \gamma_a)(11C_A-2N_f)
-36\f{S_\ep^{-1}}{\Gamma(1-\ep)}\left(-\f{4\pi\mu_R^2}{s}\right)^\ep \!\!(C_a+\ep\gamma_a)^2
\\
 +& &\!\!\!\!\!\!\!\!\!\!\!\!\ep (-1)^\ep \left(\f{\mu_R^2}{s}\right)^\ep
\Big[
36\ep^2 H_a + 2 (3+5\ep)(C_a+2\ep\gamma_a)N_f+C_A(C_a+2\ep\gamma_a)(-33-67\ep+3\ep\pi^2)
\Big]
\bigg\}\,.\nn
\eeeq
Here the coefficient $H_a$ depends on the type of the incoming partons, being \cite{Catani:1998bh,Aybat:2006mz,coefH}
\beq
H_a=
\begin{cases}
C_A^2\Big(
\f{1}{2}\zeta_3+\frac{5}{12}+\frac{11 \pi ^2}{144}
\Big)
-C_A N_f\Big(
\frac{29}{27}+\frac{\pi ^2}{72}
\Big)
+ \f{1}{2}C_F N_f+\f{5}{27}\, N_f^2\,,& \text{if }a=g \vspace*{0.3cm}\\
C_F^2\Big(\!\!
-6 \zeta_3-\frac{3}{8}+\f{\pi ^2}{2}
\Big)
+C_A C_F\Big(
\f{13}{2} \zeta_3+\frac{245}{216}-\frac{23 \pi ^2}{48}
\Big)
+C_F N_f\Big(
\frac{\pi ^2}{24}-\frac{25}{108}
\Big)\,,& \text{if }a=q,{\bar q}\;. 
\end{cases}
\eeq
The corresponding contribution to NNLO arises from the product with the Born-level matrix element, i.e., $\cm^{(2)}(\cm^{(0)})^*+(\cm^{(2)})^*\cm^{(0)}$.

Combining all the second order virtual corrections we arrive at the following expression for the two-loop virtual contribution to the cross section $\hat{\sigma}_v^{(2)}$:
\beeq
\frac{d\hat{\sigma}_v^{(2)}}{dQ^2}\!\!\!&=&\!\!\!\frac{\hat{\sigma}_0}{s}\label{virtual2}
\left(\f{\as}{2\pi}\right)^2
\delta(1-z)\,\Bigg\{
\f{S_\ep^{-1}}{72\,\ep^4}\left(\f{4\pi\mu_R^2}{s}\right)^\ep
\Bigg[
 \left(\f{4\pi\mu_R^2}{s}\right)^\ep 
\f{36\, S_\ep^{-1}(C_a+\ep\gamma_a)^2\,\Gamma^2(\ep)}{\Gamma^2(2\ep)\,\Gamma^2(1-2\ep)}
   \nn\\
   &+&
\epsilon \bigg[\left(\tfrac{\mu_R^2}{s}\right)^\ep 
\f{\Gamma(2\ep)\,\Gamma(1-2\ep)}{\Gamma(4\ep)\,\Gamma(1-\ep)\,\Gamma(1-4\ep)}\nn\\
&&\times\;\; \bigg(C_A \left(3 \pi ^2 \epsilon -67 \epsilon -33\right) (C_a+2 \ep\gamma_a)
 +
 2 N_f (5 \epsilon +3) (C_a+2 \ep\gamma_a)
   +36 H_a\, \epsilon ^2\bigg)\nn\\
   &-&\f{12\,(C_a+\ep\gamma_a)\,\Gamma(\ep)}{\Gamma(2\ep)\,\Gamma(1-2\ep)}
     \bigg(-11 C_A+6\ep \f{\hat{\sigma}^{(1)}_{\text{fin}}(\ep)}{\hat{\sigma}_0} +2 N_f\bigg)\bigg]\Bigg]
   + \f{\hat{\sigma}^{(2)}_{\text{fin}}(\ep)}{\hat{\sigma}_0}
\Bigg\}\;,
\eeeq
where $\hat{\sigma}^{(2)}_{\text{fin}}(\ep)$ is a second order finite contribution to the cross section defined by
\beq
\f{\hat{\sigma}^{(2)}_{\text{fin}}(\ep)}{\hat{\sigma}_0}=
\int
\left[
\left(\cm^{(2)}_\text{fin}\left(\cm^{(0)}\right)^*+
\left(\cm^{(2)}_\text{fin}\right)^*\cm^{(0)}
\right)
+\big|\cm^{(1)}_\text{fin}\big|^2
\right]
d\text{PS}^{(0)}
\bigg/ \int\big|\cm^{(0)}\big|^2
d\text{PS}^{(0)}\;.\label{sigfin2}
\eeq

We have evaluated all the NNLO corrections for a process with a colourless final state. The results are valid at all orders in $\ep$. To obtain a finite result we must add Eqs.~(\ref{sigmag(1)}),~(\ref{sigmaqq}),~(\ref{sigmagg}), ~(\ref{virtual2}) and the ${\cal O}(\as^2)$ of Eq.~(\ref{gsoft}) together with the counterterms coming from mass factorization (Eq.~(\ref{renormalizado})). In this way we arrive at a closed expression for the second order coefficient function $G_{a\bar{a}}^{(2)}(z)$. Expanding the result in powers of $\ep$ and keeping terms up to ${\cal O}(\ep)$ (relevant for an eventual calculation at the next order) we obtain the following result:
\beeq
G_{a\bar{a}}^{(2)}(z)&\soft&\delta(1-z)\;\Bigg\{
C_a^2\, \frac{2\pi^4}{15}
+C_a\,C_A
\left(
\frac{607}{81}+\frac{737\pi^2}{432}-\frac{407\zeta_3}{36}-\frac{7\pi^4}{48}
\right)\nn\\
&+&C_a\, N_f
\left(\!
-\frac{82}{81}-\frac{55 \pi ^2}{216}+\frac{37 \zeta_3}{18}
\right)+
\gamma_a\,\beta_0
\frac{11\pi^3}{6}
+C_a\,\frac{2\pi^2}{3}\frac{\hat{\sigma}^{(1)}_{\text{fin}(0)}}{\hat{\sigma}_0}
+\frac{\hat{\sigma}^{(2)}_{\text{fin}(0)}}{\hat{\sigma}_0}\nn\\
&+&\ep\,\Bigg[
C_a^2\left(
\frac{80 \pi ^2 \zeta_3}{3}-256 \zeta_5
\right)
+C_a N_f \left(
\frac{305 \zeta_3}{54}-\frac{488}{243}+\frac{49 \pi ^2}{81}+\frac{77 \pi^4}{960}
\right)\nn\\
&+&C_a C_A \left(
-\frac{4087 \zeta_3}{108}+\frac{101 \pi ^2 \zeta_3}{18}-\frac{43 \zeta_5}{2}+\frac{3644}{243}-\frac{707 \pi ^2}{162}-\frac{847 \pi ^4}{1920}
\right)\nn\\
&+&C_a\left(
\frac{7 \pi ^4}{9}\gamma_a-4 \zeta_3 \frac{\hat{\sigma}^{(1)}_{\text{fin}(0)}}{\hat{\sigma}_0}+\frac{2 \pi ^2}{3}\frac{\hat{\sigma}^{(1)}_{\text{fin}(1)}}{\hat{\sigma}_0}
\right)
+C_A\,\gamma_a\left(-\f{11\zeta_3}{18}+\f{1675\pi^2}{216}-\f{25\pi^4}{72}\right)\nn\\
&+&N_f\,\gamma_a\left(\f{\zeta_3}{9}-\f{125\pi^2}{108}\right)
-\f{25\pi^2}{12}H_a
+\f{7\pi^2\gamma_a}{6}\frac{\hat{\sigma}^{(1)}_{\text{fin}(0)}}{\hat{\sigma}_0}
+\frac{\hat{\sigma}^{(2)}_{\text{fin}(1)}}{\hat{\sigma}_0}
\Bigg]
\Bigg\}\nn\\
&+&\;C_a^2
\left(64\zeta_3{\cal D}_0(z)
-\frac{16\pi^2}{3}{\cal D}_1(z)+32\,{\cal D}_3(z)
\right)\nn\\
&+&\;C_a\,C_A
\left[\left(\!
-\frac{404}{27}+\frac{22\pi^2}{9}+14\zeta_3\right){\cal D}_0(z)
+
\left(\frac{268}{9}-\frac{4\pi^2}{3}\right){\cal D}_1(z)
-\frac{44}{3}{\cal D}_2(z)
\right]\nn\\
&+&\;C_a\, N_f
\left[\left(\frac{56}{27}-\frac{4\pi^2}{9}\right){\cal D}_0(z)
-\frac{40}{9}{\cal D}_1(z)
+\frac{8}{3}{\cal D}_2(z)
\right]+
8\,C_a\,\frac{\hat{\sigma}^{(1)}_{\text{fin}(0)}}{\hat{\sigma}_0}\,{\cal D}_1(z)\nn\\\label{Gab2}
&+&\ep\,\Bigg\{
C_a^2\left(
\frac{10 \pi ^4}{9}{\cal D}_0(z)-416 \zeta_3 {\cal D}_1(z)+\frac{116 \pi ^2}{3}{\cal D}_2(z)-\frac{160}{3}{\cal D}_4(z)
\right)\nn\\
&+&C_a C_A \bigg[
\left(
\frac{176 \zeta_3}{3}-\frac{2428}{81}+\frac{469 \pi ^2}{54}-\frac{\pi
   ^4}{9}
\right){\cal D}_0(z)
+\left(
-56 \zeta_3+\frac{1616}{27}-\frac{121 \pi ^2}{9}
\right){\cal D}_1(z)\nn\\
&+&\left(
\frac{8 \pi ^2}{3}-\frac{536}{9}
\right){\cal D}_2(z)
+\f{88}{3}{\cal D}_3(z)
\bigg]\nn\\
&+&C_a N_f \bigg[
\left(
-\frac{32 \zeta_3}{3}+\frac{328}{81}-\frac{35 \pi ^2}{27}
\right){\cal D}_0(z)
+\left(
\frac{22 \pi ^2}{9}-\frac{224}{27}
\right){\cal D}_1(z)
+\f{80}{9}{\cal D}_2(z)-\f{16}{3}{\cal D}_3(z)
\bigg]\nn\\
&+&C_a\bigg[\pi^2\frac{\hat{\sigma}^{(1)}_{\text{fin}(0)}}{\hat{\sigma}_0}{\cal D}_0(z)
+\bigg(
\f{28\pi^2}{3}\gamma_a+8\frac{\hat{\sigma}^{(1)}_{\text{fin}(1)}}{\hat{\sigma}_0}
\bigg){\cal D}_1(z)
-8\frac{\hat{\sigma}^{(1)}_{\text{fin}(0)}}{\hat{\sigma}_0}{\cal D}_2(z)
\bigg]
\Bigg\}+{\cal O}(\ep^2)\;,
\eeeq
where for simplicity we have set $\mu_F=\mu_R=Q$.
The dependence of the above expression on a particular process is embodied in the finite virtual contributions $\hat{\sigma}^{(1)}_{\text{fin}(i)}$ and $\hat{\sigma}^{(2)}_{\text{fin}(i)}$, the latter being defined by
\beq
\hat{\sigma}^{(2)}_{\text{fin}}(\ep)=\sum_{i=0}^\infty \hat{\sigma}^{(2)}_{\text{fin}(i)}\,\ep^i\;.
\eeq
The coefficients $C_a$, $\gamma_a$ and $H_a$ only depend on the type of the incoming partons.

The expression of Eq.~(\ref{Gab2}) coincides with the existing results for colourless final state NNLO cross sections, i.e., Drell-Yan  process \cite{Matsuura:1988sm} and  Higgs boson production via gluon fusion \cite{Catani:2001ic,Harlander:2001is} \footnote{Higher order soft (non-virtual) corrections to the ${\cal D}_i$ terms are obtained in \cite{Moch:2005ky,Ravindran:2005vv,Ravindran:2006bu}}. 
For the latter, when we consider the Higgs-gluon effective coupling in the large top mass ($M_t$) limit, the corresponding vertex factor $C_1$ has the following perturbative expansion in powers of $\as$ \cite{Kramer:1996iq,coefC1 1}:
\beq
\label{c1higgs}
C_1=
1+\f{11}{4}\f{\as}{\pi}+\left(\frac{\as}{\pi}\right)^2
   \left[\frac{2777}{288}+\frac{19 }{16}\ln\f{\mu_R^2}{M_t^2}+N_f\left(-\frac{67}{96}+\frac{1}{3}\ln\f{\mu_R^2}{M_t^2}\right)
\right]+{\cal O}(\as^3)\;.
\eeq
The ``tree-level" cross section $\hat{\sigma}_0$ is proportional to $\as^2 C_1^2$ and then can be expanded in powers of $\as$, being the lowest order term proportional to $\as^2$. The NLO and NNLO corrections are then of order $\as^3$ and $\as^4$ respectively. 
To obtain the correct result when using Eqs.~(\ref{G1}) and (\ref{Gab2}) to compute Eq.~(\ref{had}), one has to include the terms in Eq.(\ref{c1higgs})
up to the desired precision in the strong coupling constant.

The results we have presented can also be trivially expressed in a more general way, by undoing the formal $d\text{PS}^{(0)}$ integrals. Then one arrives at an identical expression for the differential cross section $d^2\hat{\sigma}/dQ^2 d\text{PS}^{(0)}$, where $d\text{PS}^{(0)}$ depends on all the internal kinematical variables of the system $F$. In this way the result would be differential except on the hadronic activity which has been integrated out.

\section{Soft-virtual approximation in Mellin space}\label{sec:mellin}

The soft limit can be defined in a more natural way by working in Mellin (or $N$-moment) space, where instead of distributions in $z$ the dominant contributions are provided by continuous functions of the variable $N$.
In fact, it was shown in ref. \cite{Catani:1996yz} that large subleading terms arise when one attempts to formulate the soft-gluon resummation problem in $z$-space (as opposed to its natural formulation in $N$-space), and that these subleading terms grow factorially with the order of the perturbative expansion. As a consequence of these spurious contributions, all-order resummation cannot be systematically defined in $z$-space. In ref. \cite{Catani:2003zt} it was shown that the soft-virtual approximation at NLO and NNLO for Higgs boson production yielded better results if defined in $N$-space.

We consider the Mellin transform $\sigma_N(Q^2)$ of the hadronic cross
section $\sigma(s_H,Q^2)$. The $N$-moments with respect to $\tau=Q^2/s_H$
at fixed $Q$ are thus defined as follows:
\begin{equation}
\label{sigman}
\sigma_N(Q^2) \equiv \int_0^1 \;d\tau \;\tau^{N-1} \;\sigma(s_H,Q^2) 
\;\;.
\end{equation} 
In $N$-moment space, Eq.~(\ref{had}) takes a simple factorized form
\begin{equation}
\label{hadn}
\sigma_{N-1}(Q^2) = \hat{\sigma}_0 \;\sum_{a,b}
\; f_{a/h_1, \, N}(\mu_F^2) \; f_{b/h_2\, N}(\mu_F^2) 
\; {G}_{ab,\, N}(\as, Q^2/\mu_R^2;Q^2/\mu_F^2) \;,
\end{equation}
where we have introduced the customary $N$-moments  of the
parton distributions ($f_{a/h, \, N}$) and of the hard coefficient function
(${G}_{ab,\, N}$):
\begin{align} 
\label{pdfn}
f_{a/h, \, N}(\mu_F^2) &= \int_0^1 \;dx \;x^{N-1} \;
f_{a/h}(x,\mu_F^2) \;, \\
\label{gndef}
G_{ab,\, N} &= \int_0^1 dz \;z^{N-1} \;G_{ab}(z) \;\;.
\end{align}
Once these $N$-moments are known,
the physical cross section in $z$-space can be obtained by Mellin inversion:
\begin{align}
\sigma(s_H,Q^2) = \hat{\sigma}_0 \;\sum_{a,b} & 
\;\int_{C_{MP}-i\infty}^{C_{MP}+i\infty}
\;\frac{dN}{2\pi i} \;\left( \frac{Q^2}{s_H} \right)^{-N+1} \;
f_{a/h_1, \, N}(\mu_F^2) \; f_{b/h_2\, N}(\mu_F^2) \nonumber \\
\label{invmt}
& \times
\; {G}_{ab,\, N}(\as, Q^2/\mu_R^2;Q^2/\mu_F^2) \;,
\end{align} 
where the constant $C_{MP}$ that defines the integration contour in the $N$-plane
is on the right of all the possible singularities of the integrand, as defined in the Minimal Prescription introduced in \cite{Catani:1996yz}.

The evaluation of $G_{a\bar{a}}$ in the limit $z\to 1$ corresponds to the evaluation of the $N$-moments $G_{a\bar{a},N}$ in the limit $N\to \infty$. In the Mellin space soft-virtual approximation we drop all the terms that vanish when $N\to \infty$ and keep only constant and logarithmic ($\ln N$) contributions \footnote{Non-diagonal channels $G_{ab,N}$ with $b\neq \bar{a}$ result in corrections which are at least ${\cal O}\left(\f{1}{N}\right)$ suppressed at the partonic level}. We introduce the notation SV-$N$ to indicate this approximation, while SV-$z$ stands for the previous results obtained in the $z$ space by keeping the most divergent terms when $z\to 1$.
Then, $z$-space and $N$-space approximations generally differ by terms that are formally subleading.

The (soft-virtual approximation to the) $N$-moments $G_{a\bar{a},N}$ can be obtained again as an expression valid to all orders in $\ep$ by using the following results:
\beeq
\int_0^1 dz\; z^{N-1}\, \delta(1-z)\!\!\! &=&\!\!\! 1\;, \\
\int_0^1 dz\; z^{N-1}\, \f{1}{(1-z)^{1+a\ep}}\!\!\! &=&\!\!\!
\f{\Gamma(N)\, \Gamma(-a\ep)}{\Gamma(N-a\ep)} =
N^{a\ep} \left[\Gamma(-a\ep)+{\cal O}\left(\frac{1}{N}\right) \right]\\
\int_0^1 dz\; z^{N-1}\, {\cal D}_0(z)\!\!\! &=& \!\!\! -\ln(N)-\gamma_E+{\cal O}\left(\frac{1}{N}\right)\;,
\eeeq
since all the contributions to the function $G_{a\bar{a}}$ have one of these three dependences on the variable $z$ (the ${\cal D}_0(z)$ terms appear in the counterterms coming from mass factorization). Then, to obtain the SV-$N$ approximation we just have to replace $\delta(1-x)\to 1$ and $(1-z)^{-1-a\ep}\to N^{a\ep}\Gamma(-a\ep)$ in Eqs.~(\ref{gsoft}),~(\ref{virtual1}),~(\ref{sigmag(1)}),~(\ref{sigmaqq}),~(\ref{sigmagg}) and (\ref{virtual2}), and replace ${\cal D}_0(z)\to -\ln(N)-\gamma_E$ in the Altarelli-Parisi splitting functions.

We can also arrive at the SV-$N$ approximation from the expanded results of Eqs.~(\ref{G1}) and (\ref{Gab2}). For this, we need the $N\to \infty$ limit of the $N$-moments of the distributions ${\cal D}_i(z)$, which can be obtained for instance from \cite{Catani:2003zt,Catani:1989ne}. The results, keeping only the ${\cal O}(\ep^0)$ terms, are the following:
\beeq
\label{g1eq}
G_{a\bar{a},N}^{(1)} &=&
4\, C_a \ln^2(N) + 8\, C_a \GE \ln(N) +C_{a\bar{a}}^{(1)} + {\cal O}\left(\f{1}{N}\right)\;,\\
\label{g2eq}
G_{a\bar{a},N}^{(2)} &=&
8\, C_a^2 \ln^4(N)
+C_a\ln^3(N)\left[32\, C_a \GE + \f{44}{9}C_A - \f{8}{9}N_f \right] \nn\\
&+& C_a\ln^2(N)\bigg[C_a\left(48\GE^2+\f{16\pi^2}{3}\right)
+C_A\left(\f{134}{9}-\f{2\pi^2}{3}+\f{44\GE}{3}\right)\nn\\
&&\;\;\;\;\;-\;N_f\left(\f{20}{9}+\f{8\GE}{3}\right)
+4\,\frac{\hat{\sigma}^{(1)}_{\text{fin}(0)}}{\hat{\sigma}_0}\bigg]\nn\\
&+&C_a\ln(N)\bigg[C_a\left(32 \gamma_E ^3+\frac{32 \gamma_E  \pi ^2}{3}\right)
+C_A\left(-14 \zeta_3+\frac{404}{27}+\frac{268 \gamma_E
   }{9}+\frac{44 \gamma_E ^2}{3}-\frac{4 \gamma_E  \pi ^2}{3}\right)\nn\\
&&\;\;\;\;\;-\;N_f\left(\frac{56}{27}+\frac{40 \gamma_E }{9}+\frac{8 \gamma_E ^2}{3}\right)+8\GE\, \frac{\hat{\sigma}^{(1)}_{\text{fin}(0)}}{\hat{\sigma}_0}
\bigg]
+ C_{a\bar{a}}^{(2)} + {\cal O}\left(\f{1}{N}\right)\;,
\eeeq
where the coefficients $C_{a\bar{a}}^{(1)}$ and $C_{a\bar{a}}^{(2)}$ are independent of $N$ and take the form
\beeq
\label{c1n}
C_{a\bar{a}}^{(1)}&=&
C_a \f{4\pi^2}{3} + 4 C_a \GE^2 + \frac{\hat{\sigma}^{(1)}_{\text{fin}(0)}}{\hat{\sigma}_0}\;,\\
\label{c2n}
C_{a\bar{a}}^{(2)}&=&
C_a^2\left(
8 \gamma_E ^4+\frac{16 \gamma_E ^2 \pi ^2}{3}+\frac{8 \pi ^4}{9}
\right)\nn\\
&+&C_a C_A \left(
-\frac{55 \zeta_3}{36}-14 \gamma_E  \zeta_3+\frac{607}{81}+\frac{404 \gamma_E
   }{27}+\frac{134 \gamma_E ^2}{9}+\frac{44 \gamma_E ^3}{9}+\frac{67 \pi
   ^2}{16}-\frac{2 \gamma_E ^2 \pi ^2}{3}-\frac{37 \pi ^4}{144}
\right)\nn\\
&+&C_a N_f \left(
\frac{5 \zeta_3}{18}-\frac{82}{81}-\frac{56 \gamma_E }{27}-\frac{20 \gamma_E
   ^2}{9}-\frac{8 \gamma_E ^3}{9}-\frac{5 \pi ^2}{8}
\right)\nn\\
&+&\gamma_a \beta_0 \f{11\pi^3}{6}
+C_a \frac{\hat{\sigma}^{(1)}_{\text{fin}(0)}}{\hat{\sigma}_0}
\left(\f{4\pi^2}{3}+4\GE^2\right)+
\frac{\hat{\sigma}^{(2)}_{\text{fin}(0)}}{\hat{\sigma}_0}\;.
\eeeq

Again, the corresponding result for Higgs production using the $ggH$ effective coupling has to be treated by incorporating the corrections  described by Eq.(\ref{c1higgs}).

\section{Phenomenological Results}\label{pheno}

To evaluate the phenomenological accuracy of the SV-$z$ and SV-$N$ approximations we compare them to the exact calculation for the processes of the kind of Eq.~(\ref{proceso}) that are available up to NNLO, i.e., Drell-Yan  and Higgs boson production via gluon fusion. We compute the corresponding hadronic cross sections for proton-proton collisions at a center-of-mass energy $\sqrt{s_H}=14\,\text{TeV}$. For simplicity, in the Drell-Yan process we only consider the photon channel, since the nature of the exchanged boson does not affect the impact of the QCD corrections we want to evaluate.

To obtain the hadronic cross section we have to perform the convolution of the partonic result with the parton distribution functions. At each order, we use the corresponding MSTW2008 \cite{MSTW} parton distribution set and QCD coupling (one-loop $\as$ at LO, two-loop $\as$ at NLO and three-loop $\as$ at NNLO).

We introduce now the notation used for the different contributions to the hadronic cross section. The up to NLO and NNLO calculations are denoted by $\sigma_{\text{NLO}}(s_H)$ and $\sigma_{\text{NNLO}}(s_H)$ respectively, while $\sigma_{\text{LO}}(s_H)$ is the LO cross section.
The contribution $\sigma^{(i)}(s_H)$ is defined as the ${\cal O}(\as^i)$ correction of the fixed order calculation $\sigma_{\text{N}^i\text{LO}}(s_H)$, i.e.,
\beq
\sigma_{\text{N}^i\text{LO}}(s_H)=\sigma_{\text{LO}}(s_H)
+\dots
+\left(\f{\as}{2\pi}\right)^i\sigma^{(i)}(s_H)
\;.
\eeq
For the corresponding soft-virtual hadronic cross sections we use the same notation adding the index SV-$z$ or SV-$N$.
The cross section $\sigma_{\text{NNLO}}^{\text{SV}}(s_H)$ contains the full NLO hard cross section.

\begin{figure}
\vspace{-0.5cm}
\begin{center}
\begin{tabular}{c}
\epsfxsize=18truecm
\epsffile{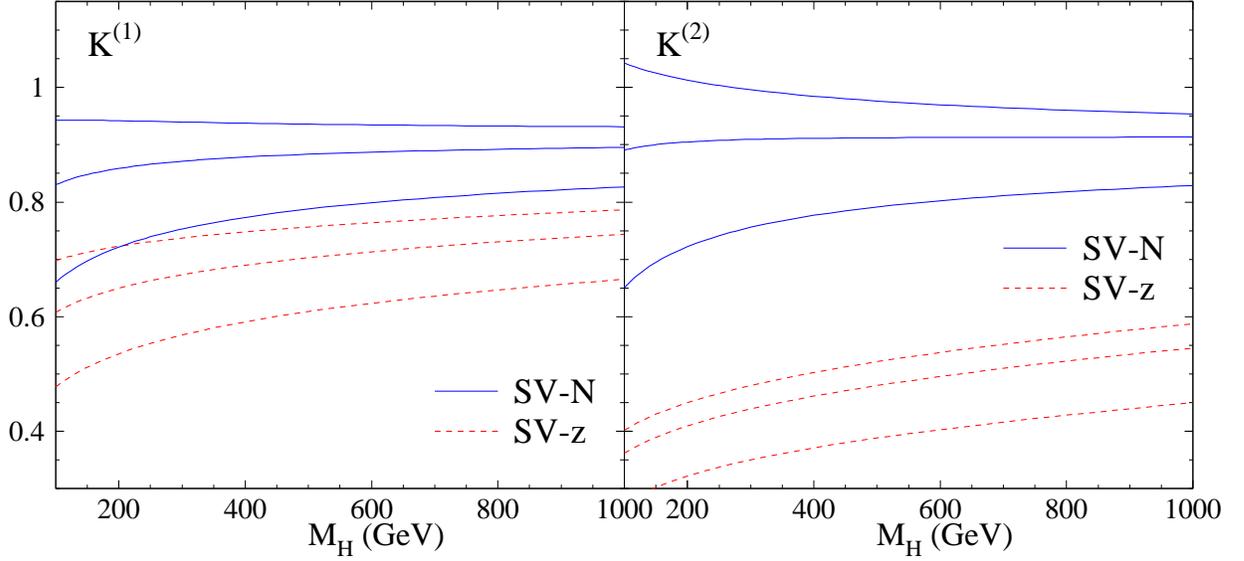}\\
\end{tabular}
\end{center}
\vspace{-1cm}
\caption{\label{K1higgs}
The ratio between SV-$N$ (solid lines) and SV-$z$ (dashed lines) approximations and the exact result for Higgs boson production $gg$ channel at the LHC ($\sqrt{s_H}=14$~TeV) for the NLO (left) and NNLO (right) contributions.}
\end{figure}

\begin{figure}
\vspace{-0.5cm}
\begin{center}
\begin{tabular}{c}
\epsfxsize=18truecm
\epsffile{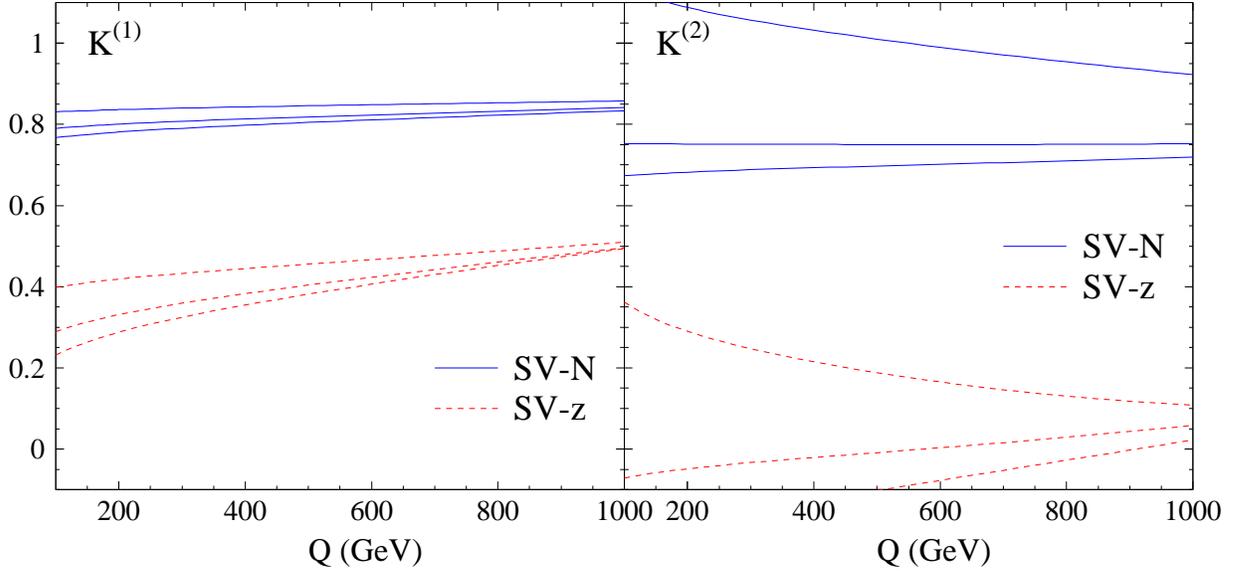}\\
\end{tabular}
\end{center}
\vspace{-1cm}
\caption{\label{K1dy}
The ratio between SV-$N$ (solid lines) and SV-$z$ (dashed lines) approximations and the exact result for the Drell-Yan $q\bar{q}$ channel at the LHC ($\sqrt{s_H}=14$~TeV) for the NLO (left) and NNLO (right) contributions.}
\end{figure}

Since the soft-virtual cross section approximates only the dominant parton subprocess, to evaluate its accuracy we first have to compare it to the partial contributions for that partonic channel. In figure \ref{K1higgs} we plot  the quantities
\beq
K_{\text{SV-}z}^{(i)}=\f{\sigma^{(i)}_{\text{SV-}z}(s_H)}{\sigma^{(i)}_{a\bar{a}}(s_H)}\;,
\;\;\;\;\;\;\;\;\;\;\;\;
K_{\text{SV-}N}^{(i)}=\f{\sigma^{(i)}_{\text{SV-}N}(s_H)}{\sigma^{(i)}_{a\bar{a}}(s_H)}\;,
\eeq
for $i=1,\,2$.
In the left-hand side of figure \ref{K1higgs} we show $K_{\text{SV-}z}^{(1)}$ and $K_{\text{SV-}N}^{(1)}$ for Higgs boson production as a function of the Higgs mass $M_H=Q$. The corresponding comparison at the next order $K_{\text{SV-}z}^{(2)}$ and $K_{\text{SV-}N}^{(2)}$ is shown in the right-hand side of the same figure.  We present the equivalent plots for the Drell-Yan process in figure \ref{K1dy}, as a function of the lepton pair mass $Q$. In all  figures the central curves are obtained by fixing $\mu_R=\mu_F=Q$, and the bands  by varying simultaneously the renormalization and factorization scales to $\mu_R=\mu_F=Q/2$ and $\mu_R=\mu_F=2Q$.

We can see from the plots that in all the cases the $N$-space approximation is in very good agreement with the exact NLO and NNLO calculations, being much more accurate than the $z$-space one, as expected since the soft-virtual approximation is defined in a more natural way in Mellin space. For Higgs boson production the SV-$N$ approximation reproduces about $90\%$ of the exact result at each order, while for the Drell-Yan process the agreement reaches the level of $80\%$. This difference is expected because the gluon distribution function grows faster than quark distributions for small  fractions of $x$, enhancing the threshold contribution to the cross section for gluon fusion processes.

\begin{figure}
\vspace{-0.5cm}
\begin{center}
\begin{tabular}{c}
\epsfxsize=18truecm
\epsffile{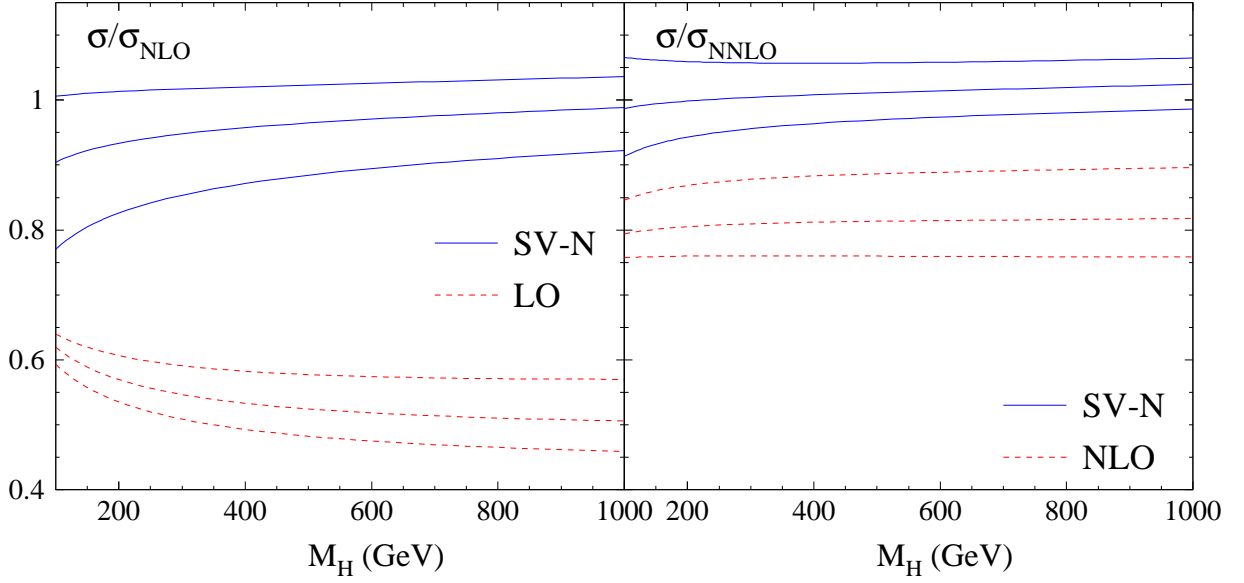}\\
\end{tabular}
\end{center}
\vspace{-1cm}
\caption{\label{fighiggs}
The ratio between $\sigma_{\text{N}^i\text{LO}}^{\text{SV-}N}$ (solid lines) and $\sigma_{\text{N}^{i-1}\text{LO}}$ (dashed lines) and $\sigma_{\text{N}^i\text{LO}}$ for Higgs boson production at the LHC ($\sqrt{s_H}=14$~TeV), for $i=1$ (left) and $i=2$ (right).}
\end{figure}

\begin{figure}
\vspace{-0.5cm}
\begin{center}
\begin{tabular}{c}
\epsfxsize=18truecm
\epsffile{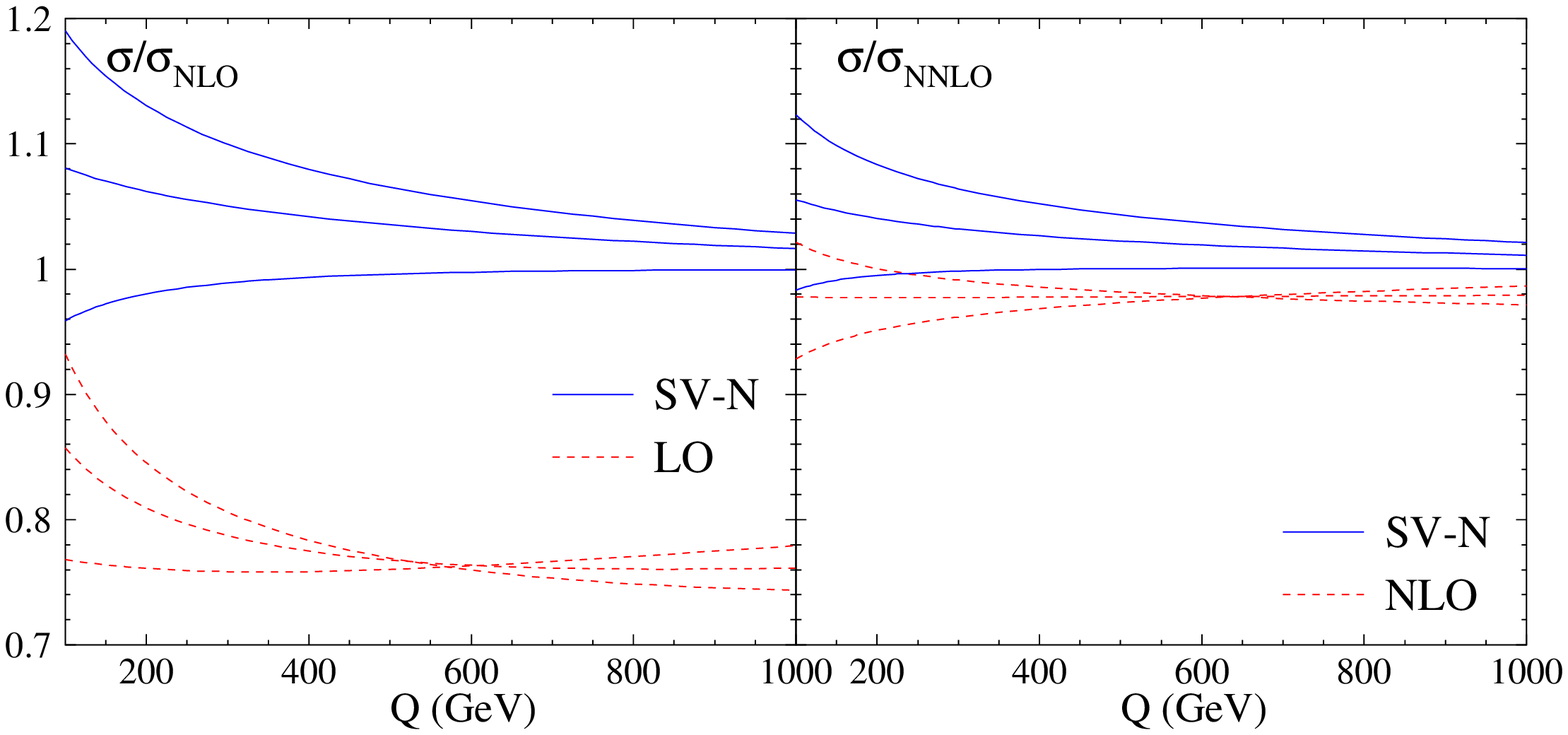}\\
\end{tabular}
\end{center}
\vspace{-1cm}
\caption{\label{figdy}
The ratio between $\sigma_{\text{N}^i\text{LO}}^{\text{SV-}N}$ (solid lines) and $\sigma_{\text{N}^{i-1}\text{LO}}$ (dashed lines) and $\sigma_{\text{N}^i\text{LO}}$ for the Drell-Yan process at the LHC ($\sqrt{s_H}=14$~TeV), for $i=1$ (left) and $i=2$ (right).}
\end{figure}

Our next step is to evaluate if the inclusion of the $\text{N}^{i}\text{LO}$ soft-virtual corrections results in an improvement over the $\text{N}^{i-1}\text{LO}$ calculation. We plot the ratio between $\sigma_{\text{N}^i\text{LO}}^{\text{SV-}N}(s_H)$ and $\sigma_{\text{N}^i\text{LO}}(s_H)$, and compare it with the ratio between $\sigma_{\text{N}^{i-1}\text{LO}}(s_H)$ and $\sigma_{\text{N}^i\text{LO}}(s_H)$, for $i=1,\,2$. This is shown for Higgs boson production in  figure \ref{fighiggs}, and for the Drell-Yan process in  figure \ref{figdy}.

For  Higgs boson production we see that the ratio between the SV-$N$ approximation and the full calculation is very close to $1$ for both NLO and NNLO \cite{Catani:2003zt}. 
We can also see that, due to the large perturbative corrections, the $\text{N}^{i-1}\text{LO}$ calculation is far from the $\text{N}^{i}\text{LO}$ one (for $i=1,\,2$). If the full NNLO result were not available, it would be clearly convenient to include the SV-$N$ approximation contribution as an attempt to improve the accuracy of the calculation. Given the fact that one can reach this conclusion 
based on the  relevance on soft-gluon emission in gluon initiated processes, it seems very reasonable to conjecture that the soft-virtual approximation might be equally accurate for any gluon-gluon fusion dominated reaction.

The situation is different for the Drell-Yan process, mainly for two reasons. In the first place, the total cross section has more sizeable contributions  from other partonic subprocesses besides the quark-antiquark one, mainly from the quark-gluon channel which is not included in the SV approximation. In the second place, the perturbative corrections are small compared to those for Higgs boson production, and then the $\text{N}^{i-1}\text{LO}$ calculation is closer to the $\text{N}^{i}\text{LO}$ one. However, we can see that the NLO SV-$N$ approximation is far more accurate than the LO result, and the NNLO SV-$N$ one is as accurate as the NLO result, improving with the invariant mass. Therefore, one might expect even more accurate results for quark initiated process where heavier states (closer to the production threshold for the same collider energy) are produced in the final state, as in double gauge boson production.

Another process for which NNLO have been recently computed is diphoton production \cite{Catani:2011qz}. One can in principle apply the formulae developed in this paper in order to compute the SV contribution. 
This is a clear counter-example for which SV corrections are not dominant and, therefore, the SV approximation fails to reproduce the main features
for the process. The main issue here is that while the Born contribution is initiated by $q\bar{q}$ annihilation, the higher order corrections from the $qg$ 
channel completely overwhelm  the Born result simply because of the large non-perturbative quark-gluon luminosity at hadronic colliders \footnote{In some cases even the formally NNLO contribution from the $gg$ channel turns out to be comparable to the Born result for the same reason.}.
This effect is particularly stressed by the implementation of asymmetric cuts in the transverse momentum of the two photons which tends to diminish the
relevance of Born like kinematic configuration and emphasise the contribution from hard (i.e, non-soft) contributions in higher order corrections \cite{Catani:2011qz}.
For some observables, like those relevant for Higgs searches, the contributions from the non-diagonal $qg$ channel amounts a few times the one arising from $q\bar{q}$ annihilation.

Therefore, even if in some kinematical conditions the SV approximation provides an accurate description of the contribution only from the $q\bar{q}$  channel, it clearly fails to reproduce the full NLO or NNLO results by a large amount.  In conclusions, the SV approximation can only be accurate enough when it is not affected by the {\it non-perturbative} enhancement of formally hard contributions from channels opening beyond the Born level with a very large partonic luminosity.

We want to remark that this issue affects basically some processes that are initiated at LO by $q\bar{q}$ annihilation, since those initiated by $gg$ are already enhanced by largest partonic luminosity at hadronic colliders. In any case, since the effect of the opening of a new channel can be appreciated at NLO, it can be checked whether such large new contribution spoils the SV approximation before attempting to use it at NNLO accuracy.

\section{Threshold Resummation}\label{resum}

Given that soft-virtual terms provide the bulk of the corrections for the processes under study, it is possible to improve over the 
 state of the art fixed-order predictions by performing soft-gluon resummation.
 In this section we consider the all-order perturbative summation of enhanced
threshold (soft and virtual) contributions to the partonic cross section. We refrain from analyzing the phenomenological impact of soft-gluon resummation and, instead, concentrate on the extraction of a {\it universal} expression for the coefficients needed to achieve next-to-next-to-leading logarithmic (NNLL) accuracy. 

The formalism to systematically perform soft-gluon resummation for 
hadronic processes,
in which a colourless massive system $F$ is produced by $q{\bar q}$ annihilation 
or $gg$ fusion, was set up in 
Refs.~\cite{Catani:1989ne,Sterman:1986aj,Catani:1990rp}.

The resummation of soft-gluon effects is achieved  by organizing the partonic coefficient function  in Mellin space as
\begin{align}
\label{resfdelta}
G_{{a\bar{a}},\, N}^{{\rm (res)}}(\as(\mu_R^2), M_H^2/\mu_R^2;M_H^2/\mu_F^2) 
&=  
\tilde{C}_{a\bar{a}}(\as(\mu^2_R),M_H^2/\mu^2_R;M_H^2/\mu_F^2) \nn \\ 
&\cdot  \Delta_{N}(\as(\mu^2_R),M_H^2/\mu^2_R;M_H^2/\mu_F^2) +
{\cal O}(1/N)\; , 
\end{align}
The large logarithmic corrections (that appear as $\as^n\ln^{2n-k} N$ in Mellin space) are exponentiated in the  Sudakov radiative factor $\Delta_N$, which depends only on the dynamics of soft gluon emission from the initial state partons.
It can be expanded as
\begin{align} 
\label{calgnnll} 
~\vspace{-.5cm} \Delta_{N}  \!\left(\as(\mu^2_R),\ln N;\frac{M_H^2}{\mu^2_R}, 
\frac{M_H^2}{\mu_F^2}\right) &=
\ln N \; g_a^{(1)}(\b0 \as(\mu^2_R) \ln N) + 
g_a^{(2)}(\b0 \as(\mu^2_R) \ln N, M_H^2/\mu^2_R;M_H^2/\mu_F^2 )
\nonumber 
\\ 
&+ \as(\mu^2_R) 
\;g_a^{(3)}(\b0 \as(\mu^2_R)\ln N,M_H^2/\mu^2_R;M_H^2/\mu_F^2 ) 
\nonumber \\
&+ \sum_{n=4}^{+\infty} \left[ \as(\mu^2_R)\right]^{n-2}
\; g_a^{(n)}(\b0 \as(\mu^2_R)\ln N,M_H^2/\mu^2_R;M_H^2/\mu_F^2 )\;. 
\end{align}

The function
$\ln N \; g_a^{(1)}$ resums all the {\em leading} logarithmic (LL) contributions
$\as^n \ln^{n+1}N$, $g_a^{(2)}$ contains the {\em next-to-leading} logarithmic 
(NLL) terms $\as^n \ln^{n}N$, $\as g_a^{(3)}$ collects
the {\em next-to-next-to-leading} logarithmic (NNLL) terms 
$\as^{n+1} \ln^{n}N$, and so forth. All the perturbative coefficients required
to construct the $g_a^{(1)}, g_a^{(2)}$ and $g_a^{(3)}$ functions are known and
only depend on the nature of the initiating partons.
Their explicit expression can be found, for instance, in Ref.\cite{Catani:2003zt,Vogt:2000ci}.

On the other hand, the function $\tilde{C}_{a\bar{a}}(\as)$ contains all the contributions that are 
constant in the large-$N$ limit. They are produced by the hard virtual 
contributions and non-logarithmic soft corrections, and  
can be computed as a power series expansion in $\as$: 
\begin{equation}
\label{Cfun}
\tilde{C}_{a\bar{a}}(\as(\mu^2_R),M_H^2/\mu^2_R;M_H^2/\mu_F^2) =  
1 + \sum_{n=1}^{+\infty} \;  
\left( \frac{\as(\mu^2_R)}{2\pi} \right)^n \; 
\tilde{C}_{a\bar{a}}^{(n)}(M_H^2/\mu^2_R;M_H^2/\mu_F^2) \;\;.
\end{equation}

The $\tilde{C}_{a\bar{a}}^{(i)}$ coefficient required to perform the resummation up to N$^i$LL can be obtained  from the corresponding fixed order computation to N$^i$LO accuracy. No general expression was known up to now for the hard coefficient, and its obtention demanded an individual computation for each process.
We show here that, given the results for the soft-virtual approximation in Mellin space presented in the previous section, it is possible to obtain the {\it universal} expressions for the $\tilde{C}_{a\bar{a}}^{(i)}$ coefficients up to NNLL accuracy. 

A direct comparison of the expansion of Eq.(\ref{resfdelta}) up to order $\as^2$ to the expression presented in Eqs.(\ref{g1eq}) and (\ref{g2eq}) shows a complete agreement for the logarithmically enhanced terms (which are fully predicted by the resummed expression), and allows the extraction of the $\tilde{C}_{a\bar{a}}^{(i)}$ coefficients from the $N-$independent terms. As a matter of fact, since it is customary to define 
 the functions $g_a^{(n)}$ such that
$g_a^{(n)}(\b0 \as \ln N)=0$ when $\as=0$, the Sudakov radiative factor can only produce
logarithmically enhanced terms (i.e., no constant terms) when expanded to any fixed order in the strong coupling constant and, therefore, the $\tilde{C}_{a\bar{a}}^{(i)}$ are exactly given by the $N-$independent terms of the soft-virtual coefficients $G_{a\bar{a},\, N}^{(i)}$  at each order in perturbation theory. Up to NNLL accuracy they are expressed by Eqs.(\ref{c1n}) and
(\ref{c2n}) as
\begin{align} 
\label{master} 
\tilde{C}_{a\bar{a}}^{(1)}&= C_{a\bar{a}}^{(1)}\;,  \nonumber \\
\tilde{C}_{a\bar{a}}^{(2)}&= C_{a\bar{a}}^{(2)}\;.
\end{align} 
The coefficients in Eq.(\ref{master}) depend, as expected, on the virtual corrections to the corresponding scattering amplitudes and on the colour factors of the initial state partons.  
We have explicitly checked that the known coefficient for Higgs \footnote{After properly taking into account the corrections to the effective $ggH$ vertex} production \cite{Catani:2003zt} agrees with the result from Eq.(\ref{master}).

The new {\it universal} expression in Eq.(\ref{master}) plus the present knowledge of the Sudakov radiative factor allows us to perform the resummation of soft-gluon emission relevant at the partonic threshold up to NNLL accuracy for any process of the kind $h_1+h_2\longrightarrow F + X $.

\section{Conclusions}\label{conc}

In this paper we have computed the NNLO soft and virtual QCD corrections for the partonic cross section of colourless-final state processes in hadronic collisions. We presented a {\it universal} expression for the corresponding cross section, whose only dependence on the process enters through the (finite part of) one- and two-loop amplitudes.

We evaluated the accuracy of the soft-virtual approximation for known processes as Drell-Yan and Higgs boson production in hadronic colliders.
We conclude that the approximation is excellent for gluon-gluon fusion initiated processes and quite accurate for quark-antiquark initiated ones for high invariant masses. For processes for which it is still not possible to compute the full NNLO corrections, counting with the corresponding soft-virtual approximation results in a clear improvement over the accuracy of the available calculation.

Finally, profiting from the soft-virtual calculation, we provided a {\it universal} expression for the coefficient needed to perform threshold resummation up to NNLL accuracy.

With the recent calculation of the three-loop quark and gluon form-factors \cite{Gehrmann:2010ue,Lee:2010cga,Baikov:2009bg}, it should be possible to attempt for an evaluation of the dominant soft-virtual corrections at N$^3$LO for a number of interesting processes (as Drell-Yan and Higgs boson production) following the approach described in this paper. 
As a first step in this direction, we have provided the necessary ingredients for ultraviolet and infrared factorization by presenting explicit results for the cross section valid to all orders in $\ep$.
The other contributions needed to obtain the N$^3$LO soft-virtual result are the following:

\begin{enumerate}

\item The soft gluon emission from a two-loop amplitude, which can be obtained from the factorization formula derived in \cite{Badger:2004uk}. The corresponding phase space integration is trivial following the same procedure indicated here for the tree-level and one-loop soft gluon emission.

\item Double soft (gluon and quark-antiquark) emission at one-loop level. The soft limit of the corresponding amplitudes may be derived following the path developed in \cite{Catani:2000pi}. The phase space integration should not be more difficult than the (double real) tree-level one described in this manuscript.

\item The tree-level triple soft emission. The corresponding soft limit of the tree-level amplitudes are accesible by either using the soft insertion rules of \cite{Catani:1999ss} or the recursion relations in \cite{Berends:1988zn}. The main complexity
of this contribution arises from the integration over the three-particle phase space, where one can forsee the extension of the method presented in \cite{Anastasiou:2012kq} to one order higher.

\end{enumerate}

\section*{Acknowledgements}

This work was supported in part by UBACYT, CONICET, ANPCyT and the Research Executive Agency (REA) of the European Union under the Grant Agreement number PITN-GA-2010-264564 (LHCPhenoNet).

\end{document}